\documentclass[12pt,preprint]{aastex}

\newcommand{\coone}{$\rm ^{12}CO(1-0)$}
\newcommand{\cotwo}{$\rm ^{12}CO(2-1)$}

\newcommand{\nhico}{$X_{CO}$}

\newcommand{\ea}{et al.}

\newcommand{\kms}{\>{\rm km}\,{\rm s}^{-1}}
\newcommand{\yr}{\>{\rm yr}}
\newcommand{\gyr}{\>{\rm Gyr}}
\newcommand{\myr}{\>{\rm Myr}}
\newcommand{\mg}{\>{\rm mag}}
\newcommand{\pc}{\>{\rm pc}}
\newcommand{\kpc}{\>{\rm kpc}}
\newcommand{\mpc}{\>{\rm Mpc}}

\newcommand{\jy}{\>{\rm Jy}}

\newcommand{\msun}{\>{\rm M_{\odot}}}

\newcommand{\dg}{^{\circ}}

\newcommand{\as}{^{\prime\prime}}
\newcommand{\am}{^{\prime}}
\newcommand{\bdm}{\begin{displaymath}}
\newcommand{\edm}{\end{displaymath}}
\newcommand{\beq}{\begin{equation}}
\newcommand{\eeq}{\end{equation}}
\newcommand{\bit}{\begin{itemize}}
\newcommand{\eit}{\end{itemize}}
\newcommand{\ben}{\begin{enumerate}}
\newcommand{\een}{\end{enumerate}}
\newcommand{\bfi}{\begin{figure}[htb]}
\newcommand{\bpfi}{\begin{figure}[p]}

\newcommand{\av}{\rm A_V}
\newcommand{\mv}{\rm M_V}
\newcommand{\htwo}{$\rm H_2$}

\newcommand{\ha}{$\rm H\alpha$}
\newcommand{\paa}{$\rm Pa\alpha$}

\slugcomment{}

\shorttitle{CO in the Nucleus of NGC\,6946}
\shortauthors{Schinnerer et al.}

\begin{document}




\def\xx{$^{[xx]}$}

\def\tbd#1{{\baselineskip=9pt\medskip\hrule{\small\tt #1}
\smallskip\hrule\medskip}}

\def\com#1{{\baselineskip=9pt\medskip\hrule{\small\sl #1}
\smallskip\hrule\medskip}}

\title{Molecular Gas Dynamics in NGC\,6946: a Bar-driven Nuclear Starburst ``Caught
in the Act''\altaffilmark{1}}

\author{Eva Schinnerer}
\affil{Max-Planck-Institut f\"ur Astronomie, K\"onigstuhl 17, D-69117 Heidelberg, Germany}
\email{schinner@mpia.de}

\author{Torsten B\"oker}
\affil{European Space Agency, Dept. RSSD, Keplerlaan 1, 2200 AG Noordwijk, Netherlands}
\email{tboeker@rssd.esa.int}

\author{Eric Emsellem}
\affil{CRAL-Observatoire, 9 avenue Charles Andr\'e, 69231 Saint Genis Laval, France}
\email{emsellem@obs.univ-lyon1.fr}

\and
\author{Ute Lisenfeld}
\affil{Dept. F\' isica Te\' orica y del Cosmos, Facultad de Ciencias,
Universidad de Granada, 18071 Granada, Spain
and
Instituto de Astrof\' isica de Andaluc\' ia, CSIC, Apdo. 3004, 18080
Granada, Spain}
\email{ute@ugr.es}

\altaffiltext{1}{Based on observations carried out with the IRAM
  Plateau de Bure Interferometer. IRAM is supported by INSU/CNRS
  (France), MPG (Germany) and IGN (Spain).}

\begin{abstract}       
  We present high angular resolution ($\sim\,1\as$ and $0.6\as$)
  mm-interferometric observations of the \coone\ and \cotwo\ line
  emission in the central $300\pc$ of the late-type spiral galaxy
  NGC\,6946. The data, obtained with the IRAM Plateau de Bure
  Interferometer (PdBI), allow the first detection of a molecular gas
  spiral in the inner $\sim\,10\as$ ($270\pc$) with a large
  concentration of molecular gas ($M_{H_2} \sim\,1.6\times10^7\msun$)
  within the inner $60\pc$. This nuclear clump shows evidence for a
  ring-like geometry with a radius of $\sim\,10\pc$ as inferred from
  the p-v diagrams. Both the distribution of the molecular gas as well
  as its kinematics can be well explained by the influence of an inner
  stellar bar of about $400\pc$ length. A qualitative model of the
  expected gas flow shows that streaming motions along the leading
  sides of this bar are a plausible explanation for the high nuclear
  gas density. Thus, NGC\,6946 is a prime example of molecular gas
  kinematics being driven by a small-scale, secondary stellar bar.
\end{abstract}

\section{Introduction}
A large fraction of nearby spiral galaxies experience intense star formation 
within a few hundred parsec from their nuclei. Many detailed studies 
exist of the stellar populations produced in these starbursts and the 
physical conditions of the ionized gas surrounding them. These studies
mostly use imaging and/or spectroscopy at optical and near-infrared (NIR)
wavelengths which can provide sub-arcsecond resolution and thus allow the
study of star formation processes at scales of individual giant molecular
clouds (GMCs). Case studies of this type include NGC\,1808
\citep*{kra94,tac96}, IC\,342 \citep*{boe97}, NGC\,253 \citep{eng98}, and also
the subject of this paper, NGC\,6946 \citep{eng96}.

A necessary, although maybe not sufficient, condition for such nuclear 
star formation is the inward transport of molecular gas -- the raw material 
for producing stars -- from the outer galaxy disk.
If efficient enough, this transport can, over time, lead to a
re-distribution of the baryonic mass, and thus can change the
appearance of a galaxy, in the sense that the (stellar) mass
distribution becomes more centrally concentrated. Recently, such 
secular evolution processes have received renewed attention,
because observational evidence has accumulated that at least at the
late-type end of the Hubble sequence, galaxy bulges grow and modify
their appearance well after their host disks have formed
\citep[for a detailed review see][]{kor04}.
Without a detailed understanding of the mechanisms that govern the 
transport of disk gas into the innermost regions of disk galaxies, 
our understanding of galaxy evolution therefore remains incomplete.

Over the past decades, much progress has been made in the theoretical
understanding of the mechanisms that cause gas rotating inside the
galaxy disk to lose angular momentum and to radially drift towards the
nucleus. Bars and spirals have often been mentioned as potential
drivers for inward gas motion, but the efficiency of such processes is
rather badly understood, and certainly depends on the physical state
of the gas (e.g.  viscosity) and on the detailed characteristics of
the acting perturbation \citep[e.g. the pattern speed of the
bar][]{com85}. \cite{ath92} has performed a whole set of
hydro-dynamical simulations exploring the influence of the main
parameters of the bar: central concentration, axial ratio, quadrapole
moment, and pattern speed. At the same time,
\cite{com93} have explored the behavior and appearance of bars in
early and late type spirals showing that quite different types of bars
are expected for these two classes of galaxies. Direct comparison of
models to observations have shown that gas flow induced by stellar
bars can, for example, explain the complex kinematics in our Galactic
Center \citep{bin91} or the gas flow in the double-barred spiral
galaxy M100 where the two nested bars need to be dynamically decoupled
\citep{gar98}. Recently, (hydro-)dynamical modeling of the gas flow in
central kiloparsec by e.g. \cite{eng00,mac02,mac04a,mac04b} has
provided detailed predictions about the properties of the gas in the
presence of single or nested bars. In addition, the influence of bars
on the gas dynamics has been well-tested by observations on spatial
scales of a few kpc, for example in NGC\,4303 \citep{sch02}, NGC\,5383
\citep{she00}, or NGC\,5005 \citep{sak00}.

However, relatively little is known about whether similar mechanisms
are also at work to produce the high (molecular) gas density required
to enable the intense {\bf nuclear} star formation (i.e. inside
$\sim\,100\pc$ from the dynamical center) observed in many spirals.
While it is tempting to invoke small-scale inner bars and dynamical
resonances in a way similar to the standard model of kpc-scale
starbursts, there are few observational tests of these models on the
scales of nuclear starbursts. The reason lies mostly in the fact that
maps of the molecular gas can rarely match the angular resolution
obtained at optical/NIR wavelengths. For example, the most
comprehensive interferometric CO survey of nearby galaxies, the BIMA
SoNG \citep{reg01,hel03}, has on average a spatial resolution of
$6\as$, or $300\pc$ at a distance of $10\mpc$. The IRAM PdBI key
project NUGA is reaching angular resolution below $1\as$: however it
is focused on a smaller sample of nearby spiral galaxies containing an
active nucleus \citep{gar03}. Therefore, it is only in a few nearby
(and bright) galaxies that one can hope to obtain a more detailed
picture of the molecular gas dynamics within and around the nuclear
starbursts, and to observationally constrain dynamical models on
scales of a few tens of parsecs.

NGC\,6946 provides such an opportunity. At a distance\footnote{The
  exact distance to NGC\,6946 is somewhat uncertain. The Nearby
  Galaxies Catalog of \cite{tul88} reports a distance of $5.5\mpc$,
  but more recent estimates are $5.9\,\pm\,0.4\mpc$ to the center of
  the NGC\,6946 group based on the luminosity of blue supergiants in 7
  satellite galaxies \citep*{kar00}, and $5.7\,\pm\,0.7\mpc$ based on
  the expanding photo-spheres of Type II supernovae \citep{sch94}.
  Throughout this paper, we adopt the \cite{tul88} value because it
  simplifies comparison to most other papers in the literature, and is
  consistent with the more recent estimates.} of $5.5\mpc$
\citep[$1\as\,=\,27\pc$][]{tul88}, it is one of the nearest large
spiral galaxies outside the local group, and is currently undergoing
intense nuclear star formation \citep[e.g.][]{eng96}. The enhanced
star formation activity is unlikely to be triggered by a dynamical
interaction, since HI observations of the neighborhood of NGC\,6946
\citep{pis00} show neither obvious tidal streams nor any companion
galaxies that are massive and close enough to have any dynamical
impact on NGC\,6946. However, the HI velocity field is not symmetric
at the outskirts of the disk \citep[e.g.][]{cro02}. NGC\,6946 is
classified as an SAB in the RC3 catalog \citep{vau91}; a stellar bar
with a length of $\ge 120\as (3.5\kpc)$ and position angle of $\rm
17^o$ has been identified in NIR images \citep{reg95,men06}. In
addition, and more importantly for the data presented in this paper,
NGC\,6946 harbors a small inner bar with an estimated major axis
diameter of $\sim\,200\pc$ which is most clearly visible in the NIR
images of \cite{elm98} and \cite{kna03}. \cite{reg95} have argued that
the inner bar is too small to be responsible for the pile-up of
molecular gas in the center of NGC\,6946. However, its role for the
gas dynamics inside the central $\sim\,100\pc$ could not be studied in
previous CO observations due to their limited spatial resolution.

In this paper, we present new $0.6\as$ resolution CO maps that reveal
in detail the molecular gas flow around (and within) the inner bar.
We will use these data, in combination with dynamical models, to
demonstrate that NGC\,6946 is in fact a showcase for a nuclear
starburst that is sustained by bar-driven inflow of molecular gas. We
describe our observations and the data reduction procedure in
\S\,\ref{sec:data}, and present the observational results in
\S\,\ref{sec:results}. We will compare these results in
\S\,\ref{sec:flows} to dynamical models of gas flows within the
stellar potential (derived from NIR data), and will show that the
observed gas morphology is well explained by the model. The relation
between the nuclear star formation and the molecular gas is discussed
in \S\,\ref{sec:center}. In \S \ref{sec:comp} we conclude with a
discussion of the importance of similar processes for the nuclear
starbursts and the evolution of galactic nuclei. The results are
summarized in \S \ref{sec:summary}.

\section{Observations and Data Reduction}\label{sec:data}

\subsection{The CO Data}\label{subsec:codata}

The \coone\ and \cotwo\ lines at 3mm and 1mm, respectively, were
simultaneously observed at two days in February 2002 using the IRAM
Plateau de Bure interferometer (PdBI) with 6 antennas in its A and B
configurations providing 30 baselines ranging in length from 30 to 400
m. The resulting uv coverage is shown in Fig. \ref{fig:10channels} and
\ref{fig:21channels}. The resulting resolution using uniform
weighting\footnote{In the convention of the IRAM GILDAS package,
uniform weighting corresponds roughly to a robust=0 parameter in the
AIPS package.} is $1.25\as\,\times\,1.01\as$ (PA $22\dg$) and
$0.58\as\,\times\,0.48\as$ (PA $-165\dg$) for the \coone\ and \cotwo\
line, respectively. The calibration and mapping were done in the
standard IRAM GILDAS software packages CLIC and MAPPING
\citep{gui00}. The phase center of the observations was at
20h34m52.36s +60d09m15.96s in the observed epoch of J2000.0. All
velocities are observed relative to $\rm
v_{ref}(LSR)\,=\,65\,\kms$. The integration times for NGC\,6946 and
the phase calibrator 2037+511 were divided into 20 and 3 scans with
lengths of 60 and 45 seconds, respectively. During the observations
the typical system temperature in the sideband containing the line of
interest at 115 (230) GHz was about 270 (300\,--\,350)\,K.
 
For the bandpass, phase, and amplitude calibration we used
observations of the quasar 2037+511 which was observed about every 20
minutes. The flux calibration was also checked on this source and
resulted in an upper limit for the uncertainty in the flux calibration
of about 10(17)\% at 115(230)\,GHz. The pixel scales of the data cubes
are 0.2$\as$/pixel and 0.12$\as$/pixel at 3mm and 1mm, respectively,
with an image size of 512$\times$512 pixels to ensure that the dirty
beam is well-sampled while all emission (even outside the primary
beam) can be CLEANed. Channel maps were CLEANed with up to 1000
iterations down to a flux limit of 1.5$\sigma$. In order for the CLEAN
process to converge, we needed to define CLEAN regions. For the
\coone\ data cube this was achieved using a 0th moment map derived from
a CLEANed cube without any set CLEAN areas. Thus we defined the same
CLEAN region for all channels. For the \cotwo\ data, it was necessary
to separately define CLEAN regions for each individual channel. The
correct choice of CLEAN regions was verified by inspecting the
resulting residual maps for the presence of significant structure. The
lack of missing information on very low spatial frequencies in the uv
plane (i.e. very short baselines) is likely the main reason why a
single region file was not applicable.

The spectral resolution during the observations was 1.25 MHz per
channel both at 3mm and 1mm. The final rms in the binned 6$\kms$ wide
channels is 4 and 12 mJy/beam for the \coone\ and \cotwo\ line,
respectively, as measured in the channels where line emission is
present. Using the equation to estimate the rms noise provided by
IRAM\footnote{e.g. {\tt
http://www.ram.fr/GENERAL/calls/calls\_s06/node34.html}} a theoretical
rms of 3.5 and 8 mJy is expected for natural weighting under standard
observing conditions and for our chosen bandwidth.  Taking into
account that we expect a higher rms due to the uniform weighting, the
agreement between theoretical and actual rms is good.

The moment maps were calculated using the GIPSY task 'MOMENTS'
requiring that emission is present above a certain clipping level in
at least two adjacent channels. This minimizes the inclusion of noise
peaks in the maps. We used a clipping level of 3$\sigma$ for the 0th
moment (intensity map), and a 4$\sigma$ level for the 1st and 2nd
moments (velocity field and dispersion map). We present the resulting
channel maps of the \coone\ and \cotwo\ line emission in Fig.
\ref{fig:10channels} and \ref{fig:21channels}, respectively. Spectra
extracted from the central $4.8\as\,\times\,4.8\as$ are shown in Fig.
\ref{fig:spec}. All data are presented without a primary beam
correction applied.

\subsection{Archival Data}\label{subsec:arch}

In order to determine the missing flux in our data, we retrieved 
the \coone\ data of the BIMA SoNG project \citep{hel03} from NED. 
The data cube contains a combination of BIMA and 12m NRAO
observations, and should contain the entire \coone\ flux, since the short 
spacings missed by the BIMA mm-interferometer are provided via the 12m 
single-dish data. The data cube has an angular resolution of 
6.0$\as \times$4.9$\as$, a channel width of 10$\kms$, and an rms noise 
level of 61 mJy/beam \citep[for details see][]{hel03}. 
We calculated moment maps in a similar way to the PdBI data
using a 3$\sigma$ cut-off for all three moments.

For comparison to our CO maps, we also obtained a number of optical
and NIR images from various archives. For these, we used an H-band
image from the 2MASS survey \citep{jar03} as a common reference frame
by aligning field stars. We used the K$_s$ band image of \cite{kna03}
to construct the gravitational potential for our qualitative dynamical
model (see \S \ref{subsec:model}) of the inner stellar bar.
In addition, we used the I-band continuum and H$\alpha$ line
emission images of \cite{lar99} to compare the CO morphology to the
large-scale star formation. For the
modeling in \S\,\ref{subsec:model}, we also retrieved from the CDS
archive the HI and H$\alpha$ velocity curves from \cite{car90} and
\cite{blais04}, respectively.

Finally, we obtained higher resolution, space-based images from the
{\it Hubble Space Telescope} (HST) archive for a more detailed
comparison with the nuclear CO morphology: (i) an I-band image taken
with WFPC2 in the F814W filter, (ii) a NICMOS H-band image, and (iii)
a continuum-subtracted Pa$\alpha$ image. The latter two images were
taken as part of the SINGS legacy project (HST-ID 9360, PI: R.
Kennicutt). All HST images were also aligned to the 2MASS image using
field stars. Given the resolution difference between ground- and
space-based observations and the limited number of stars available for
alignment, we estimate that our absolute astrometry of the HST images
is good to within $\sim\,0.3\as$ relative to the 2MASS image.

\section{Observational Results}\label{sec:results}

\subsection{CO Morphology: a Nuclear Gas Spiral}\label{subsec:morph}

In Fig.~\ref{fig:co-sum}, we present the intensity maps for both CO
lines. The overall \coone\ morphology resembles that of the lower
resolution maps (of $\sim\,4\as$ and $\sim\,6\as$) of the BIMA and NMA
arrays \citep{reg95,hel03,ish90}. The two molecular arms extending to
the north and south are clearly seen in the PdBI \coone\ intensity
map.  There is some evidence for multiple spiral arms, but whether
this is a true $m=4$ spiral structure as claimed by \cite{reg95} is
not obvious.

The molecular gas concentration in the inner
$15\as$ ($400\pc$) could not be resolved with previous 
\coone\ observations: \cite{ish90} proposed that it is a nuclear
molecular bar based on its elongation and the presence of non-circular
motions. On the other hand, \cite{reg95} concluded that the
concentration is formed by the continuation of the outer (stellar)
spiral arms. Recently, \cite{mei04} suggested that the central
concentration is actually a barely resolved ring which is fed via the
outer northern and southern gas lanes which are produced by the
large-scale bar.

Our new CO observations resolve for the first time the central
molecular gas concentration into a S-shaped spiral with an
extent of about $10\as$ ($270\pc$). For the remainder of this paper,
we will use the term ``nuclear spiral'' to describe this feature. The
higher-resolution \cotwo\ map (Fig. \ref{fig:co-sum}b) shows in
detail how the molecular gas is distributed within the nuclear
spiral. The gas is concentrated in three bright clumps (indicated in
Fig. \ref{fig:feat}), one on either side of the nucleus at a radial
distance of $\sim\,4\as$ or $120\pc$, and a third clump that coincides
with the dynamical center, at least within the uncertainty of our
astrometry (see \S\,\ref{subsec:rot}).

Overall, the morphology of the nuclear gas spiral closely resembles a
scaled-down version of the gas flow in a large-scale stellar bar
predicted by dynamical models \citep[e.g.][]{ath92}. The two outer clumps are
connected to the nuclear one via straight gas lanes which run along a
position angle of $PA \sim\,-40^{\circ}$. Both gas lanes are offset
from each other by about $2\as$ perpendicular to their length. Also,
the three clumps are extended perpendicularly to the length of the gas
lanes. In \S\,\ref{subsec:model}, we will show that the nuclear spiral
can indeed be well explained by the gas responding to the gravitational
potential of the small inner stellar bar in the center of NGC\,6946.

\subsection{CO Kinematics: Radial Gas Inflow}\label{subsec:kinematics}

Overall, the \coone\ velocity field (Fig. \ref{fig:co-vel}a) exhibits a
well-defined pattern. The dominant motion is from North-East to
South-West, in agreement with a position angle $PA \sim\,242^{\circ}$
of the kinematic major axis derived from HI observations
\citep{cro02}. One prominent exception is the gas in the northern gas
lane which seems to move from East to West. The zero velocity curve
reveals several changes (or wiggles) in the region of the nuclear
spiral structure. The changes are closely associated with the spiral
structure as is clearly seen in the \cotwo\ data (Fig.
\ref{fig:co-vel}b). Along the inner ridge of the southern clump the
velocity of the gas is blue-shifted with respect to the systemic
velocity, whereas it appears red-shifted along the inner ridge of the
northern clump. Assuming that the spiral arms are trailing, the
northeastern side of NGC\,6946 is the near side. This is consistent
with the high extinction northeast of the nucleus that is apparent in
optical and NIR images which is expected to occur on the near side of
galaxies \citep{vau58}. The observed velocities therefore suggest that
the molecular gas motion has a radial component towards the nuclear
clump. Inside the nuclear clump ($r \le 2\as$), the position angle of
the major kinematic axis is closer to an east-west orientation. All
this indicates the presence of strong non-circular gas motions, most
likely inwards streaming onto the nuclear clump which reduces the
apparent circular velocity at a given radius.

The position-velocity (p-v) diagrams (Fig.\,\ref{fig:pv}) along the
major and minor kinematic axes of $242^{\circ}$ and $152^{\circ}$
\citep{cro02} can be used to derive additional kinematic and geometric
properties of the nuclear gas clump itself, i.e. inside $30\pc$ from
the kinematic center:

{\it First,} there are two peaks in the steep part of the p-v diagrams
along the major kinematic axis at $r \sim\,0.45\as$
(Fig.~\ref{fig:pv}a,c) which suggests that the nuclear clump is not a
smooth disk, but rather has a hole inside, i.e. is
ring-like. \cite{sak99} called this a ``hole feature'' in the case of
a steeply rising rotation curve with circular motion only (see their
appendix A for a detailed discussion). Alternatively, such a feature
could be caused by Keplerian motion due to a central black hole or
higher velocity dispersion in the center. The second scenario is
likely not the case for NGC\,6946, as velocity dispersion map shows no
such increase towards the dynamical center (see below). In addition,
two peaks are also seen along the minor axis (Fig. \ref{fig:pv}b,d) at
$\rm r\,\sim\,0.33\as$. The spatial offsets indicate a radius for this
ring-like distribution (hereafter: nuclear ring) of about $\rm
r_{ring}\,\approx\,(9\,-\,13)\pc$ when taking inclination into
account. It is interesting to note that a third peak is present in the
\cotwo\ p-v diagram which is not seen in the \coone\ one. This might
be either due to angular resolution effects or it might indicate real
variations in the line ratios of the molecular gas in this putative
nuclear ring.
 
{\it Second,} as already noted by several authors
\citep{ish90,reg95,mei04}, the non-zero velocity along the minor axis
strongly suggests streaming motions in the central 300\,pc. Our new CO
data allow us to trace this non-zero component well inside the nuclear
clump. Inside a radius of $\sim\,1\as$ (see Fig. \ref{fig:pv}d) and
coincident with the location of the nuclear ring at a radius of $r
\sim\,0.45\as$, the gas changes the sign of its velocity along the
minor axis suggesting that strong streaming motion are associated with
this nuclear ring at a distance of only about $11\pc$. It is not
clear, if this non-circular motion could be only mimicking the
presence of a ring-like structure. However, the twin-peaked structure
of the central clump suggests that a true depression is present.

{\it Third,} the nuclear flux distribution appears to be asymmetric
with respect to the observed reference velocity. The p-v diagrams in
Fig.\,\ref{fig:pv} are plotted relative to the reference velocity of
our PdBI observations which was set to $v_{ref}(LSR)=65\kms$, based on
the CO p-v diagram in Fig. 4 of \cite{sak99}. The obvious offset of
the p-v structure with respect to zero velocity then indicates that
the gas in the nuclear ring moves with about $-25\kms$ relative to the
reference velocity implying a systemic velocity of $\rm
v_{sys}(LSR)\,\approx\,40\kms$. However, a slightly higher value of
$\rm v_{sys}(LSR)\,\approx\,50\kms$ (indicated by the horizontal line
in the p-v diagrams of Fig.\,\ref{fig:pv}) for the systemic velocity
is suggested when taking into account the motion of the bulk of the
emission. A systemic velocity of $\rm v_{sys}(LSR)\,=\,50\kms$ has
also been suggested by several authors based on the extended atomic
gas emission \citep[e.g.][]{car90,cro02}. Thus, we are led to conclude
that the nuclear flux distribution is not symmetric with respect to
the systemic velocity and has an offset relative to $\rm v_{sys}$ by
about $-15\kms$.

The velocity dispersion in the disk outside the nuclear spiral
structure is generally below $10\kms$ in both CO lines. The \coone\
dispersion map (Fig. \ref{fig:co-disp}a) shows values between about 25
and 35\,$\kms$ inside the spiral structure, whereas the derived
velocity dispersion in the \cotwo\ map (Fig. \ref{fig:co-disp}b) is
about 5\,$\kms$ lower. This difference is very likely due to the
higher spatial resolution of the \cotwo\ data which reduces beam
smearing effects. One interesting aspect is that the velocity
dispersion seems to increase when crossing the southeastern gas lane
from the leading to the trailing side, i.e. from northeast to
southwest. A similar pattern is only barely indicated for the
northwestern gas lane. Such a behavior is expected when streaming
motions due to a stellar bar are present. This will be further
explored in \S \ref{subsec:model}.
 
The velocity dispersion inside the nuclear clump is about $10-20\kms$
higher than throughout the nuclear spiral. The dispersion reaches a
maximum of about $50$ and $42\kms$ in the \coone\ and \cotwo\ line,
respectively. This dispersion peak is slightly offset -- as already
expected from the p-v diagrams -- from both the intensity peak and the
dynamical center (see \S\,\ref{subsec:rot}). Given our angular
resolution of 0.6$\as$, it is difficult to decide how much of this
offset is an artifact produced by beam smearing effects.

\subsection{The Dynamical Center and the CO Rotation Curve}\label{subsec:rot}

We used two approaches to derive a rotation curve from our \coone\ and
\cotwo\ data: a) an automated fit to the velocity fields using the
GIPSY task 'ROTCUR', and b) interactive fitting of the 3-dimensional
data-cube using the GIPSY task 'INSPECTOR'. Both tasks are based on
the assumption that the gas motions can be described by a model of
rotating tilted rings. In principle, the inclination, position angle,
center, and systemic and circular velocity of the individual rings are
free parameters. In order to constrain this large parameter space, we
fixed the inclination and the position angle of the rings to the
values found for the large-scale HI disk of $i \sim\,38^{\circ}$ and
$PA \sim\,242^{\circ}$ \citep{car90,cro02}. We assume that these
describe best the parameters of the galactic disk in NGC\,6946 and
that the central CO emission is in the plane of the galaxy. 

The dynamical center was determined from the \cotwo\ p-v diagrams in
Fig.\,\ref{fig:pv} as the position which maximizes the symmetry of the
velocity structure relative to the systemic velocity. As discussed in
\S\,\ref{subsec:kinematics}, we adopt a systemic velocity of
$v_{sys}(LSR)=50\kms$ (i.e. $15\kms$ lower than the reference velocity of
the observations). This approach yields the following coordinates for
the dynamical center: R.A.: 20h34m52.355s, Dec.: +60d09m14.58s
(J2000). We estimate the uncertainty of our dynamical center position
to be $\sim\,0.3\as$, dominated by the rather complex velocity
structure in the central $<1\as$. Within this uncertainty, our
coordinates agree well with those measured by \cite{mei04}, see also
Fig. \ref{fig:center}.

After ``freezing'' the dynamical center to the above coordinates, the
final rotation curves were obtained with only the circular velocity as
a free parameter. The derived (deprojected) rotation curves for the
\coone\ and \cotwo\ line all show a steep rise at small radii and then
level off at about $125\kms$ (Fig.\,\ref{fig:rot}). The steeper
gradient of the \cotwo\ rotation curve at $r<6\as$ is expected because
of the higher angular resolution of these line data. The rotation
curves obtained with the 'INSPECTOR' task are, in general, steeper
than the ones produced with the 'ROTCUR' task. This is also expected
because they result from direct fits to the 3-dimensional data cubes,
and are therefore less affected by beam smearing \citep{swa99}.

In order to test whether the drop of the PdBI CO rotation curves
beyond a radius of $\sim\,10\as$ is real or an artifact due to the
sampling of our data, we used the BIMA SoNG \coone\ data which have
lower angular resolution ($\sim\,6\as$), but cover a wider field. The
BIMA SoNG rotation curve over-plotted in Fig.\,\ref{fig:rot} was
derived using 'ROTCUR' with the same parameter set as for the PdBI
data. As expected from the larger BIMA beam, it shows an inner rise
which is shallower. At larger radii, the rotation curve stays flat out
to $75\as$. The apparent drop for $r>4\as$ ($r>9\as$) seen in the PdBI
\cotwo\ (\coone ) curves produced with 'ROTCUR' is an artifact
produced by a lack of spatial coverage in our data, emphasized by
the fact that 'ROTCUR' uses only 2-dimensional velocity fields instead
of the full data cube. Our CO rotation curves are consistent with that
of \cite{mei04} once beam smearing effects are taken into account.

\subsection{Molecular Gas and Dynamical Mass}\label{subsec:mass}

The amount of molecular gas contained within the central few
hundred parsec provides a gauge for the efficiency and sustainability 
of the current nuclear star formation activity. In order to estimate 
the molecular gas mass, we have calculated (after
correcting for the primary beam response) the \coone\ flux
contained in five distinct apertures in the center of NGC\,6946: 
i) the central 50$\as$, i.e. the area containing the entire 
\coone\ emission in Fig.~\ref{fig:co-sum}a, ii) the nuclear spiral, i.e
the entire area of Fig.~\ref{fig:co-sum}b, and iii) the
three bright CO clumps within the nuclear spiral that are marked 
in Fig.~\ref{fig:feat}. The results are summarized in Table~\ref{tab:COfluxes}.

The total flux contained in the \coone\ map of
Fig.~\ref{fig:co-sum}a is about 50\% of that measured over the same
area in the short spacing-corrected BIMA SoNG data (see Tab.
\ref{tab:COfluxes}). This means that about half the emission stems
from an extended component which is not sampled by the PdBI baselines.
As for components ii) and iii), missing flux is less of a concern
because their spatial extent is well-matched to the angular scales
probed by the PdBI observations.

In order to convert the measured CO flux into a molecular gas mass for
the individual components (see Tab. \ref{tab:COfluxes}), we used $\rm
M_{H_2}[M_{\odot}] = \frac{S_{CO} \times c^2}{2 k \nu^2} X_{CO} \times
D^2 \times 2 m_p = 519295 \times S_{CO}[Jy\,km\,s^{-1}] D[Mpc]^2
X_{CO} / \nu[Hz]^2$ \citep[e.g.][]{bra01} with the Distance $\rm D$,
speed of light $\rm c$, Boltzmann constant $\rm k$, and the proton
mass $\rm m_p$. We assumed a \nhico\ conversion factor of $\rm X_{CO}
= 2 \times 10^{20} cm^{-2}\,(K\,km\,s^{-1})^{-1}$ \citep{str88} and an
observed frequency of $\rm \nu = 115.246\,GHz$. This
results in a molecular gas mass of $3.1\times10^8\msun$ contained in
the entire map, and $9.3\times10^7\msun$ within the nuclear spiral.
About 1/6 of this latter mass is located within the nuclear clump
which has a size of about $2\as\,\times 1.5\as$ ($\approx 60\pc \times
45\pc$). The northern clump is about half as massive as the nuclear
clump, while the southern clump contains about 50\% more molecular
gas.  Note, however, that the value of the \nhico\ conversion factor
in the center of NGC\,6946 has been estimated to be 4-5 times lower
than the standard value \citep{mei04,wal02}. The gas masses listed in
Table~\ref{tab:COfluxes} could therefore be overestimated by this
factor.

Assuming that solid body rotation is a reasonable approximation for
the gas dynamics in the central $4\as$, we can estimate the dynamical
mass contained in the inner $110\pc$. From the \cotwo\ rotation curve
(Fig. \ref{fig:rot}), we find a rotational velocity of $v \approx
155\kms$ at a radius of $r = 2\as$ ($\approx 55\pc$) which translates
into a dynamical mass of $M_{dyn}\sim\,3.1\times\,10^8\msun$. For
comparison, \cite{eng96} give a $2\sigma$ upper limit on the dynamical mass
of $M_{dyn} \sim\,3\times10^8\msun$ inside a radius of $4.25\as$
($\approx 115\pc$) based on the velocity dispersion of stellar
absorption lines as measured from long slit NIR spectra with $R\approx
3000$. 

After accounting for the contribution of He (36\%) to the total gas
mass, the mass in the central $4\as$ of $\sim\,3.1\times\,10^7\msun$
is between about 2\% (for the lower limit in the $\rm X_{CO}$
conversion factor) and 10\% of the enclosed dynamical mass. This value
is in reasonable agreement with the molecular gas fraction of
$\sim$\,5\% for the inner $6\as$ found by
\cite{mei04}. Therefore, the molecular gas fraction in the center
of NGC\,6946 appears to be similar to the molecular gas fraction
observed in the central kiloparsec of nearby spiral galaxies
\citep{sak99,she05}.

\section{What Drives the Gas Flow?}\label{sec:flows}

\subsection{The Large Scale}\label{subsec:outer}

The classification of dynamical modes in NGC\,6946 is hampered by the
fact that the spiral structure is complex and can not be easily
classified. \cite{elm92} find evidence for a superposition of $m=2$
and $m=3$ modes. From the analysis of a combined HI and CO (residual)
velocity field, \cite{cro02} also finds evidence for an outer $m=3$
and an inner $m=2$ mode. Both modes appear to have the same pattern
speed. The derived corotation radius is at $(4.5\pm0.5)\am$ or
$\approx\,(7.3\pm0.8)\kpc$.

NGC\,6946 is classified as a barred spiral both in the RC3 catalog and
LEDA \citep{vau92,pat03}. The length of the large-scale bar or oval
was first reported by \cite{reg95} to be about $120\as$
($\approx\,3.3\kpc$), well inside the corotation radius of the spiral
structure
\citep[see also][]{elm98}. Recently, \cite{men06} have confirmed this
length from 2MASS images, and give a position angle of $\rm \sim\,17^o$
for the bar. The flat abundance gradient throughout the central
$5\kpc$ of NGC\,6946 reported by \cite{bel92} could be explained in the 
context of the presence of a bar, which is expected to enhance radial 
mixing.

However, the large-scale bar appears to leave no obvious signature in
the observed gas kinematics, which might not come as a surprise given
its low ellipticity of 0.15 \citep{elm98}. The BIMA SoNG observations
\citep{hel03}, which cover most of the area of the large-scale bar,
show a concentration of molecular gas in the inner $0.5\am$, with some
hints of gas lanes extending toward the north and south. The higher
resolution PdBI observations of the \coone\ line resolve the inner gas
concentration, but there is no obvious pattern which can readily be
associated with this large-scale bar. On the other hand, the
comparison between the \coone\ intensity distribution and the \ha\
line emission in the inner arcminute is not inconsistent with
gas responding to a stellar bar. As can be seen in
Fig.\,\ref{fig:rgb}, the \ha\ emission is strongest on the leading
side of the CO lanes, especially on the northern side. This spatial
correlation between molecular gas and current star formation seems
fairly typical as it is observed for many large-scale stellar bars
\citep{she02}.

Nevertheless, we conclude that the details of the
{\bf nuclear} gas distribution and kinematics studied in this paper 
are likely not affected by the presence of the large-scale bar, and
we will therefore ignore its presence for our modeling efforts described
in \S\,\ref{subsec:model}.

\subsection{The Inner NIR Bar}\label{subsec:bar}

NGC\,6946 is known to host a small-scale stellar bar \citep{reg95}
which has been studied in more detail by \cite{elm98}. In
Fig.\,\ref{fig:bar}, we compare the NIR morphology from the K$_s$ band
image of \cite{kna03} to the \cotwo\ distribution in the central kpc
of NGC\,6946. The apparent diameter of the inner bar has been estimated by
\cite{elm98} from ellipse fitting to be $\sim\,8\as$ with an
ellipticity of 0.4 and a position angle of about 140\degr. Deriving
the exact length of the inner bar is not straight-forward in the case
of NGC\,6946, as large extinction is present in the inner kpc (see
\S \ref{sec:center}).
 
Therefore, the $8\as$ length of the inner bar reported by \cite{elm98}
should probably be regarded as a lower limit. This is supported by
Fig.\,\ref{fig:bar} which compares the observed isophotes of the K$_s$
band emission with (theoretical) circular isophotes of $5\as$,
$10\as$, and $15\as$ radius projected into the galaxy plane: only the
outermost K$_s$ isophote with a radius of $15\as$ appears close to
being circular. For smaller radii, the deviation from axisymmetry is
still high, strongly suggesting that the NIR bar is more extended than
the $\sim\,8\as$ diameter reported by \cite{elm98}.  A larger NIR bar
size is also supported by gas dynamical models which show that gas
lanes always form inside the bar potential \citep[e.g.][]{ath92}.
Assuming an ellipticity of 0.4 for the inner bar \citep{elm98}, its
apparent length should therefore be $\sim\,16\as$ \citep[i.e. twice as large
as measured by][]{elm98} in order to encompass the gas lanes.

As shown in \S \ref{subsec:morph}, the morphology of the CO spiral in
NGC\,6946 is remarkably reminiscent of gas lanes in a barred
potential. Such a distribution can only be stationary, if either (a)
the rotation is rigid in the central few arcseconds over the extent of
the structure or (b) it is a density wave (caused by e.g. the NIR
bar). As the velocity is not linearly increasing with radius outside a
radius of $\sim\,3.5\as$ (see Fig. \ref{fig:rot}), this leaves only a
density wave as the cause for the spiral structure unless the spiral
is observed at a special time.

As already described in \S \ref{subsec:kinematics}, the velocity field
shows various signatures expected for gas moving in a barred
potential. The iso-velocity lines are twisted away from the kinematic
minor axis in the region of the inner NIR bar. There is also an abrupt
change in the orientation of the iso-velocity contours in the
innermost 2$\as$, with the zero velocity curve almost along a
North-South direction. Such a behavior is expected in the presence of
an Inner Lindblad Resonance (ILR) where the outer $x_1$ orbits
elongated along the bar major axis are replaced by $x_2$ orbits inside
the ILR which are elongated along the bar minor axis
\citep[e.g.][]{ath92}.

Figure \ref{fig:pvbar} shows four representative p-v diagrams along cuts
that run perpendicular to the northern and southern gas lanes 
(see Fig.\,\ref{fig:center} for the exact location of the p-v cuts). 
An abrupt change in velocity is apparent in the southern
p-v diagrams, and - less prominently - also in the northern
ones. Such a jump in velocity is expected to occur as the gas enters
the shock region inside the gas lane \citep[see for comparison Fig. 11
in][]{ath92}.

The qualitative arguments presented in this section suggest that the
observed CO morphology and kinematics can be explained by the gas
responding to a barred potential. We now proceed to the next level of
detail, and construct a simple dynamical model of NGC\,6946 in an
attempt to qualitatively reproduce the observed central CO
distribution and kinematics in the context of gas flows within an
inner small bar.

\subsection{Modeling the Central Kiloparsec}\label{subsec:model}

Our modeling efforts follow the formalism used in \cite*{egf03},
\citep[see also][]{wad01}: an initial axisymmetric mass model is
perturbed via the addition of a bar-like structure, and gas orbits are
then derived from a treatment of the epicycle approximation including
a damping term.

The first step in creating an axisymmetric mass model is to obtain a
three-dimensional, axisymmetric {\it luminosity} model which
reasonably approximates the observed photometry. For this purpose, we
use the K$_s$ band image of NGC~6946 obtained by \cite{kna03}. We
first deproject the K$_s$ band image assuming that the galaxy is
two-dimensional with its major axis at a position angle of
PA$=242\degr$ and an inclination of i=38\degr. We then obtain an
analytical face-on model by simply fitting the resulting deprojected
image with a Multi-Gaussian-Expansion model \citep[MGE
hereafter]{mbe92, emb94} constraining all components to be circular
(i.e. an axis ratio equal to 1) and taking into account a Gaussian
point spread function of FWHM$ = 1\as$ to account for atmospheric
seeing. The resulting MGE model is arbitrarily thickened with an axis
ratio varying continuously from 0.25 in the outer part to 0.8 in the
central $2\as$. Figure~\ref{fig:massmodel} compares this model after
projection (using again the assumed inclination of 38\degr) with the
original K$_s$ band image. The fit is reasonably good considering the
imposed constraint of axisymmetry. The vertical height of the
luminosity distribution is rather unconstrained, and our model is
obviously not designed to reproduce the large-scale stellar bar and
the spiral structure. It will therefore only serve, as mentioned
above, as a reference model for the inner few arcseconds of the galaxy
in order to probe the potential role of the inner tumbling bar
potential. We have, however, checked that the results described below
are not significantly influenced by, e.g., a change in the vertical
height. In the following, we also assume that the large-scale bar does
not play any major role in the central 10$\as$, as discussed in
\S\,\ref{subsec:outer}.

The knowledge of the spatial luminosity distribution allows us to
derive the corresponding axisymmetric gravitational potential by
simply using Poisson equation and assuming a constant mass-to-light
ratio $M/L_{Ks}$ in the K$_s$ band. We constrain $M/L_{Ks}$ by
comparing the predicted circular rotation curve $V_c(r)$ in the
equatorial plane with the observed deprojected HI and H$\alpha$
rotation curves (Fig. \ref{fig:Vcmodel}) obtained by \cite{car90} and
\cite{blais04}, respectively: we obtain $M/L_{Ks} = (0.72 \pm 0.1)$, a
value adopted for the rest of this Section.

We then add a bar-like perturbation $\delta \Phi$ to the axisymmetric
gravitational MGE potential in the
form described in \cite{egf03}, with $\delta \Phi = \Phi_2 \times
\cos{\left( 2 \theta \right)}$. This component corresponds to the
small-scale inner NIR bar. We assume for this bar component the same
$M/L_{Ks}$ ratio as for the stars in the outer galactic disk, although this
may be an overestimate considering the presence of young stars in 
the central region. Gaseous orbits are then derived via the
linear epicycle approximation, and including an artificially induced
damping (simulating the dissipative nature of the gas) described by a
parameter $\Lambda$ [$\rm km^2\,s^{-2}$]. The radius, force and pattern
speed of the bar are provided by $R_{bar}$ [$\as$], $Q_{bar}$
[$\rm km^2\,s^{-2}$], the amplitude of the perturbation, and $\Omega_p$
[$\rm km\,s^{-1}\,kpc^{-1}$], respectively. The bar radius $R_{bar}$ 
corresponds to the radius where the contribution from the bar component
is designed to drop (as $r^{-4}$).

In order to constrain some of these parameters, we assume that the bar
encompasses the southern and northern clump: this implies that the
radius of the bar is $R_{bar} > 5 \arcsec$, and imposes an upper limit
on its pattern speed $\Omega_p \sim\,\rm 1100\,km\,s^{-1}\,kpc^{-1}$,
so that the bar lies within its corotation radius. We have derived
models for $5\as \le R_{bar} \le 10\as$ and $\rm
100\,km\,s^{-1}\,kpc^{-1}$ $\le \Omega_p \le$ $\rm
1100\,km\,s^{-1}\,kpc^{-1}$, with the aim to qualitatively reproduce
the observed distribution and kinematics of the \coone\ line emission
in the central $10\as$. For each model, we compute the observed
density distribution and first two velocity moments, the velocity
field and the dispersion map. The predicted maps are derived by first
assuming an initially unperturbed radial density distribution for the
gas, and take into account the resolution of the \coone\ data with a
clean beam size of 1.12$\as$. Only models with $6.5\as \le R_{bar} \le
8\as$ and $\rm 510\,km\,s^{-1}\,kpc^{-1}$ $\le \Omega_p \le$ $\rm
680\,km\,s^{-1}\,kpc^{-1}$ can roughly reproduce the length and shape
of the \coone\ gas lanes. It is however difficult to further constrain
these parameters given the observed patchiness in the distribution of
the \coone\ line emission, and the asymmetry of its velocity field.
The derived range of bar lengths of $13\as$ to $16\as$ is consistent
with the inferred range of observed apparent bar length of about $8\as <
l_{bar,obs} < 30\as$ (when correcting for inclination: $10\as < l_{bar}
< 38\as$; see \S \ref{subsec:bar}).

A comparison between a representative model and the data is shown in
Fig.~\ref{fig:kinmodel}. The model parameters were $\Lambda =
\rm 1800\,km\,^2\,s^{-2}$, $R_{bar} = 7.5\as$, $Q_{bar} =
\rm (350)^2\,km^2\,s^{-2}$ and $\Omega_p = \rm 510\,km\,s^{-1}\,kpc^{-1}$.
The corresponding corotation radius is then at $\sim\,9.5\as$ ($\sim\,260\pc$), and the southern and northern clumps are close to the Ultra
Harmonic 4:1 Resonance (UHR) at $6.7\as$. This model reproduces the
extent and position angle of the observed straight gaseous lanes
which end near the UHR with an abrupt change in the pitch angle: such
a structure resembles the one observed in the ionized gaseous inner
component in the early-type galaxy NGC~2974 \citep{egf03}, also
interpreted in the context of an inner tumbling bar. In the present
model, the corresponding ILR lies then at $\sim\,2.2\as$, just
outside the extent of the nuclear clump.

The predicted velocity field exhibits clear deviations from circular
motions: this is best illustrated by the wiggles in the zero velocity
isovelocity line which significantly departs from the kinematic minor
axis, with a peak difference of about 30\degr\ along the gas lanes. At
a radius of about 5\arcsec, the zero velocity curve gets back to the
kinematic minor axis, a feature clearly present in the \coone\
velocity field of NGC\,6946: this may be interpreted in the model as a
change in the orbital structure near the UHR of the bar component. The
velocity twist in the central 2\arcsec\ of the model is however not
strong enough to reproduce the one observed in the \coone\ velocity
field. Overall, the predicted velocities are too high by about 20\%:
as mentioned above, this may indicate that $M/L_{Ks}$ is lower in the
central 500~pc than in the outer part of the disk due to the presence
of young stars.

The crowding of orbits near the gas lanes also implies strong shocks
and high velocity dispersion values. The \coone\ maps of NGC\,6946
indeed hint at the presence of such high dispersion values located
on the inner (trailing) side of the gas lanes.
The exact locations of the high dispersion regions along the gas lanes
relative to the densest regions strongly depend on the prescription
for the shocks (damping), which are clearly not properly addressed
with the present static model. In this model, high central dispersion
values are indeed mostly due to orbit crowding and unresolved velocity
broadening, i.e. beam smearing effects. A more realistic physical
treatment of the shock regions would require full hydro-dynamical
simulations.

The NIR isophotes and the \coone\ kinematics in the central 10$\as$ of
NGC\,6946 exhibit signatures of a stellar
bar (see \S \ref{subsec:bar}). Although the model shown here
(Fig.~\ref{fig:kinmodel}) is clearly not meant as a fit to the data, it
illustrates that the observed gas distribution and kinematics are
indeed qualitatively consistent with the presence of an inner stellar
bar of about $200\pc$ radius. Such a bar would thus naturally explain:

\begin{itemize}
\item the presence of straight gas lanes, on the leading edge of the bar, 
which are then regions of shocks located roughly between the ILR and UHR.
\item the abrupt change in the gas density distribution near the UHR, a feature
also observed in more realistic hydro-dynamical simulations, such as the one
performed by \cite{mac02}, and
\item the associated strong twist of the iso-velocities near the end 
of the gas lanes.
\end{itemize}

In this context, the gas concentration in the central clump could be
interpreted as the result of gas accumulation within the ILR of the
inner bar. Models such as the one described in \cite{mac02}, indicate
that such gas could settle on a nuclear ring: this would also be
consistent with the ring-like morphology of the nuclear clump as
inferred from its \cotwo\ p-v diagrams (see
\S\,\ref{subsec:kinematics}). The high central gas density, as well as
the presence of shocks in the inner $150\pc$, are likely necessary
ingredients for the high star formation efficiency in the center of
NGC\,6946. More realistic hydro-dynamical simulations, possibly in
combination with higher resolution maps of the molecular gas, are
required to confirm these results, and to further constrain the
mechanisms that cause the nuclear starburst in NGC\,6946.

It is interesting to compare the case of NGC\,6946 to that of the
early-type galaxy NGC\,2974. The CO distribution observed in NGC\,6946
is strikingly similar to the ionized gas structure revealed by WFPC2
narrow band images of NGC\,2974 \citep{egf03} which reveal straight
gas lanes and the presence of a central gaseous concentration within
the inner $10\pc$. These features were interpreted in NGC\,2974 as
being driven by an inner bar with a diameter of $\sim\,540\pc$ and a
pattern speed around $700\rm km\,s^{-1}\kpc^{-1}$, values comparable
to the ones derived for the inner bar of NGC\,6946. The main
difference then lies in the overall low dust and gas content
($\sim\,6.8\times10^4\msun$ of ionized gas) and the absence of
significant star formation in NGC\,2974, as expected for its E/S0
classification.  The inner bar of NGC\,2974 may therefore represent an
early-type analog of the one in NGC\,6946, the gas reservoir of
NGC\,2974 not being sufficient to have triggered a significant
starburst. In both cases, however, the dynamical time-scale is
similarly short ($\sim\,10^6\yr$ at about $200\pc$ from the center),
which implies a strongly time-varying gaseous distribution.

\section{Discussion}
\subsection{The Central Starburst}\label{sec:center}

The central $300\pc$ ($\sim\,11\as$) of NGC\,6946 currently experiences
intense star formation. This is evident from the multitude of hydrogen
recombination lines in the spectrum of its nuclear region \citep[see
e.g.][]{eng96}, but also by the morphology of the \paa\ emission in
the {\it HST/NICMOS} image shown in Fig.~\ref{fig:hstcomp}.
It is thus not surprising that NGC\,6946 is classified to have an 
HII region nucleus based on the [OI]/\ha\ and [SI]/\ha\ line 
ratios \citep*{ho97}.

In order to estimate the current star formation rate (SFR) in the
nucleus of NGC\,6946, we have measured the \paa\ flux of the central
$8\as$ from the image of \cite{boe99a} which is $\rm
5.5\,\times\,10^{-13}\,erg\,s^{-1}\,cm^{-2}$.  We assume for the
moment an average extinction of $\av\,=\,4.6\,mag$ as derived for the
southern ``hot spot'' in Fig.~\ref{fig:hstcomp} by \cite{qui01} from
estimates of the \ha /\paa\ line ratio.  Then applying the extinction
law of \cite{car89} with $\rm R_V\,=\,3.1$, and assuming an intrinsic
\ha /\paa\ line ratio of 7.41 (appropriate for case B recombination
with $\rm T_e\,=\,5000\>K$ and $\rm N_e\,=\,100\,cm^{-2}$, see
Osterbrock 1989), we can derive an intrinsic \ha\ flux of $\rm
7.6\,\times\,10^{-12}\>erg\,s^{-1}\,cm^{-2}$.  Converting to \ha\
luminosity (with $D\,=\,5.5\mpc$) and applying the relation of
\cite{ken83}, this translates into a current SFR of
$0.25\msun\yr^{-1}$.

However, this simple calculation most likely leads to an underestimate
of the true SFR due to the large extinction present in the nucleus of
NGC\,6946. The fact that the extinction in the center of NGC\,6946 is
rather high and uneven is apparent from the HST I-band image in
Fig.\,\ref{fig:hstcomp}. The stellar continuum emission is rather
asymmetric, and noticeably reduced in regions of high gas density. It
is also apparent from the J-K color map in Fig. 10 of \cite{elm98}
which shows a remarkable similarity to the CO distribution
demonstrating that the dust associated with the molecular gas causes
significant extinction. It is clear from Fig.\,\ref{fig:hstcomp} that
both line and continuum emission are significantly suppressed in
regions of high gas density, i.e. the CO spiral and especially the
three clumps marked in Fig.~\ref{fig:feat}.

One aspect of uncertainty is the geometry of the obscuring dust.  For
example, \cite{eng96} derive an extinction of $\av \sim\,10.4$ for the
central 8.5$\as$ based on emission lines assuming that stars and dust
are mixed. For comparison, they infer $\av \sim\,4.3$ assuming a
screen geometry for the dust, which agrees well with the \cite{qui01}
value. An alternative approach to obtain an extinction estimate for
these regions which are basically opaque for recombination lines is to
use the molecular gas surface density. For example, assuming that the
molecular gas in the nuclear clump is uniformly distributed over its
area, we obtain an \htwo\ surface mass density of $\rm M_{H_2}/A =
1.6\times10^7\msun/(\pi\times(27\pc)^2) \approx 7000\msun\pc^{-2}$.
This corresponds to an \htwo\ column density of $\rm N(H_2) \approx
4.4\times10^{23}\,cm^{-2}$. Adopting the Galactic extinction relation
of $\rm A_V/N_H = 6\times 10^{-22} cm^2\mg$ \citep{dra03}, we derive
an extinction in the optical of $\av = 6 \times 10^{-22} \times 2
N(H_2) = 528\mg$. Even taking into account the uncertainty in the
\nhico\ conversion factor discussed in \S\,\ref{subsec:mass} which
could lead to extinction values that are lower by a factor of 4-5, it
is clear that the {\it average} extinction across the central $\approx
60\pc$ is most likely higher than the values derived by \cite{eng96}
and \cite{qui01}. This in turn implies that the true SFR is likely
higher than the $0.25\msun\yr^{-1}$ value derived above. Observations
in the mid-infrared would be required to solve this large discrepancy.

From the high \htwo\ surface density across the nuclear gas clump, we
conclude that the true nucleus of NGC\,6946 is highly obscured even at
NIR wavelengths. It is thus possible that the two bright \paa\ peaks
on either side of the dynamical center (Fig.\,\ref{fig:hstcomp})
are merely an artifact caused by the high extinction of the central CO
clump. If, however, the peaks are true ``hot spots'' and their \paa\
emission is indeed produced by photoionization regions around young
stars (as opposed to shocks resulting from a possible outflow from the
nucleus), it is interesting to point out that their location roughly
coincides with the ILR inferred from the dynamical model in
\S\,\ref{subsec:model}. We thus might be observing a small-scale
analog to the kiloparsec-scale starburst rings seen in many barred
spirals.

Finally, we point out that the active star formation as traced by the
\paa\ emission is, in general, not well correlated with the molecular
gas. While for very young starbursts, this is a consequence of the gas
shroud obscuring the view into the starburst, once massive stars have
formed, their intense UV-radiation, stellar winds, and supernova
explosions all combine to photo-dissociate and/or expel the CO gas. As
a consequence, it is highly unlikely to observe both dense molecular
gas, and signatures of active star formation in the same location.

\subsection{Is the Center of NGC\,6946 Peculiar?}\label{sec:comp}

It is interesting to compare the properties of the inner few hundred
parsecs in NGC\,6946 to those of similar late type spiral
galaxies such as our Galaxy or IC\,342. Although NGC\,6946 is of 
somewhat later Hubble type (Scd vs. Sbc) and slightly less luminous 
($\mv=-19.1$ vs. $\mv = -20.6$) than the Milky Way, in many respects both 
galaxies turn out to be fairly similar. For example, the dynamical 
mass within the inner $\sim\,110\pc$ is nearly identical in both galaxies: 
$\sim\,3.1\times10^8\msun$ for NGC\,6946 (see \S\,\ref{subsec:mass})
and $\sim\,3.5\times10^8\msun$ for the Galactic Center \citep[derived from
Fig. 7.1 in][]{gen94}. Moreover, the distribution and kinematics of 
the molecular gas appear similar in both galaxies. This can be seen
by comparison with the recent study of \cite{saw04} who derived the 
face-on geometry of the molecular gas in the central kiloparsec of the
Milky Way from a quantitative analysis of the \coone\ line
emission and OH absorption lines. Their derived \coone\ intensity map and 
velocity field \citep[Fig. 8 in][]{saw04} show a remarkable resemblance 
to the \coone\ distribution in NGC\,6946, especially when accounting for
differences in angular resolution. In both cases, the molecular gas 
distribution is elongated with an apparent extent of $500\pc
\times 200\pc$. \cite{saw04} suggest that in the case of the Milky Way,
this gaseous ``bar'' is actually formed by two barely resolved spiral 
arms formed as a response to a large-scale stellar bar. 

However, there are also important differences. In NGC\,6946, the
molecular gas appears more concentrated than in the Milky Way.  The
gas mass $\rm M_{H_2}$ contained within the nuclear spiral, i.e.
within a radius of $140\pc$ is about $9.3\times10^7\msun$.  For the
Milky Way, \cite{lau02} have reported a gas mass of $2\times10^7\msun$
within $r \le 230\pc$, which implies a gas density about a factor of
up to 12 lower than in NGC\,6946. The high molecular gas density in
the center of NGC\,6946, in particular the presence of a massive gas
clump at the galaxy nucleus, appears somewhat unusual. In IC\,342,
another late-type spiral with enhanced star formation, the molecular
gas is located in a ring-like structure around the nuclear cluster,
but there is little gas inside the about $100\pc$ diameter ring:
$M_{H_2} \sim\,2\times10^5\msun$ in the inner $30\pc$ \citep{schi03}.
Most likely, some of these morphological differences are the
consequence of time evolution. It is obvious that the currently
observed gas distribution can be only a short-lived phenomenon, as the
fueling process itself as well as the ongoing star formation impact
the gas. 

Recently, \cite{bou05} suggested that large-scale bars in gas-rich
spiral galaxies are short-lived. In their dynamical models, the bar
almost dissolves within a few dynamical time-scales ($\sim\,1-2\gyr$ in
their model) due to the angular momentum transfer from the infalling
gas to the stellar bar. A similar behavior is expected for inner bars
where the dissolution time would be also only a few dynamical
time-scales (F. Combes, priv. comm.). Thus, the inner bar of NGC\,6946
could significantly weaken within $10^7\yr$ and thus reduce the gas
flow. In addition, the (infalling) gas might quickly be used up by the
ongoing massive star formation. This together could result in a dearth
of molecular gas in the inner few tens of parsec, resulting in a very
different appearance of NGC\,6946 in the future.

However, if the NIR bar is long-lived, a relaxation oscillator
mechanism for (nuclear) starbursts as it has been suggested for the
300\,pc ring in the Galactic Center by \cite{sta04} could also provide
a time-varying molecular gas content. This mechanism is based on the
gravitational collapse of the gas into giant gas clouds in the ILR
ring suggested by \cite{elm94}. \cite{sta04} used the properties of
the Galactic Center molecular gas derived by \cite{mar04} to estimate
a critical density of $\rm n(H_2) \sim 10^{3.5}\,cm^{-3}$ above which
the molecular gas in the ILR will become instable and
self-gravitating. The sudden formation of these clouds can result
in a starburst. In addition, these larger clouds can then be disrupted
by tidal forces and thus inflow to the center can occur. \cite{mei04}
have estimated the density of the central cloud E1 which roughly
coincides with the CO spiral to $\rm \sim\,10^{4.2}\,cm^{-3}$ showing
that a warmer, denser molecular gas is associated with the
starburst. In order to test the scenario of a 'relaxation oscillator'
for NGC\,6946 a better knowledge of the ISM and its properties are
required.

Alternatively, the gas distribution might respond strongly to the
energetic feedback from the star formation processes. The nuclear
cluster in IC\,342 experienced its last episode of star formation some
$60\myr$ ago \citep*{boe99b}. It seems plausible that the cavity in
the molecular gas distribution could be a consequence of that last
starburst, either because the intense UV-radiation has destroyed the
molecules, or because stellar winds and/or SN explosions have expelled
the gas. In contrast, NGC\,6946 is actively forming stars, and has
been doing so for only $\sim\,7\myr$ \citep{eng96}.  While it is
currently not possible to decide which of the mechanisms discussed
here more strongly affects the molecular gas distribution, they both
will reduce the amount of molecular gas in the nucleus of NGC\,6946.
One might thus speculate that in another $50\myr$ or so, the central
molecular gas distribution in NGC\,6946 will be more similar to the
one observed in IC\,342.

There is no clear evidence for the presence of an AGN in NGC\,6946,
although \cite{hol03} find two luminous X-ray sources in the vicinity
of the nucleus. It is interesting to note that within the errors, the
fainter one coincides with our best estimate of the dynamical center
(see Fig.\,\ref{fig:center}). Given that this source most likely is
subject to the rather high extinction due to the central CO clump, its
intrinsic X-ray luminosity could be well above the Eddington limit for
a neutron star, providing some tentative evidence for a central black
hole in NGC\,6946.  On the other hand, the X-ray spectrum of the
nuclear source is not substantially different from that of the rest of
the X-ray source population, and might have another explanation
related to the intense nuclear star formation.

\section{Summary and Conclusions\label{sec:summary}}

The sub-arcsecond resolution CO observations presented in this paper
resolve for the first time the molecular gas concentration within the
inner $300\pc$ of NGC\,6946. The molecular gas is located in an
S-shaped spiral structure with three high-luminosity clumps located at
the dynamical center and on either end of the spiral. Strong streaming
motions are associated with the straight gas lanes of this spiral
structure and are even observed within the inner few 10 parsec.
Supported by simulations of gas orbits in a perturbed mass
model of NGC\,6946, we conclude that the observed morphology is best
explained by the gas responding to the non-axisymmetric potential of
an inner, small-scale stellar bar.

We find that a molecular gas mass of $\rm M_{H_2}
\sim\,1.6\times10^7\msun$ has accumulated within a radius of $27\pc$.
Although spatially unresolved in our data we find evidence from
analysis of the p-v diagrams that the gas inside the nuclear clump is
likely distributed in a ring which would then lie close to the ILR of the
inner stellar bar. Such a morphology closely resembles that predicted
(and observed) for the ILR of large-scale bars. While a direct
confirmation of the nuclear gas ring has to await higher resolution
data, our analysis suggests that inner bars are indeed capable of
funneling gas towards galactic nuclei to within at least a few tens of
parsec.

Comparison of the NGC\,6946 results to similar studies for the Milky
Way and IC\,342 indicates that the amount and distribution of
molecular gas in the central 100 parsec of NGC\,6946 is fairly different and
suggests that the molecular gas properties in galactic nuclei might be
strongly variable over time. We discuss scenarios where the observed
CO properties critically depend on the
nuclear star formation history and the (exact) shape of the nuclear
stellar potential present at the time of the observations.
Our results, however, demonstrate that - at least in gas-rich spirals
- significant amounts of molecular gas can be present in the immediate
vicinity ($\rm r < 60\,pc$) of the nucleus; they are likely
transported there by an inner bar and - at least a significant
fraction - effectively converted into stars. This adds some support
to scenarios in which secular evolution plays a significant role for
the ongoing morphological evolution of spirals at the late-type end of
the Hubble sequence.

\acknowledgments 

We would like to thank J. Knapen and R. de Jong for their quick and
constructive feedback on the $\rm K_s$ band data. In addition, we are
grateful to D. Calzetti and M. Sosey for providing the NICMOS images
of NGC\,6946. We thank the anonymous referee for insightful comments
which helped to improve the paper. UL acknowledges support by the
Spanish Ministry of Education, via the research projects
AYA\,2005-07516-C02-01, ESP\,2004-06870-C02-02, and the Junta de
Andaluc\' ia. This paper has benefited from use of the LEDA database
(http://leda.univ-lyon1.fr). This research has made use of the
NASA/IPAC Extragalactic Database (NED) which is operated by the Jet
Propulsion Laboratory, California Institute of Technology, under
contract with the National Aeronautics and Space Administration.

\clearpage

\begin{deluxetable}{lrr}
\tablecaption{
$^{12}$CO\,(1-0) fluxes and Molecular Gas Masses\label{tab:COfluxes}}
\tablewidth{0pt}
\tablehead{
\colhead{Component} & 
\colhead{$\rm S_{CO}$}  &  
\colhead{$\rm M_{H_2}$}   \\
          & 
\colhead{[$\rm Jy\,km\,s^{-1}$]} & 
\colhead{[$10^7\msun$]} } \startdata
total           &    1310  &    30.98 \\
Nuclear spiral  &     394  &     9.31 \\
Nuclear Clump   &      67  &     1.58 \\
Clump North     &      31  &     0.74 \\
Clump South     &     101  &     2.38 \\
Central 4''     &      95  &     2.25 \\
\enddata 
\tablecomments{\coone\ line fluxes and molecular
  gas masses for various components of the molecular gas distribution
  derived from the primary beam corrected data. The components are
  indicated in Fig. \ref{fig:feat}. Note that the molecular gas masses
  could be lower by up to a factor of about $4-5$ due to the
  uncertainty in the conversion factor $X_{CO}$ (see text). The Helium
  fraction of 36\% has not been taken into account for the values of
  $\rm M_{H_2}$.}
\end{deluxetable}

\clearpage
\begin{figure}
\includegraphics[angle=0,scale=.8]{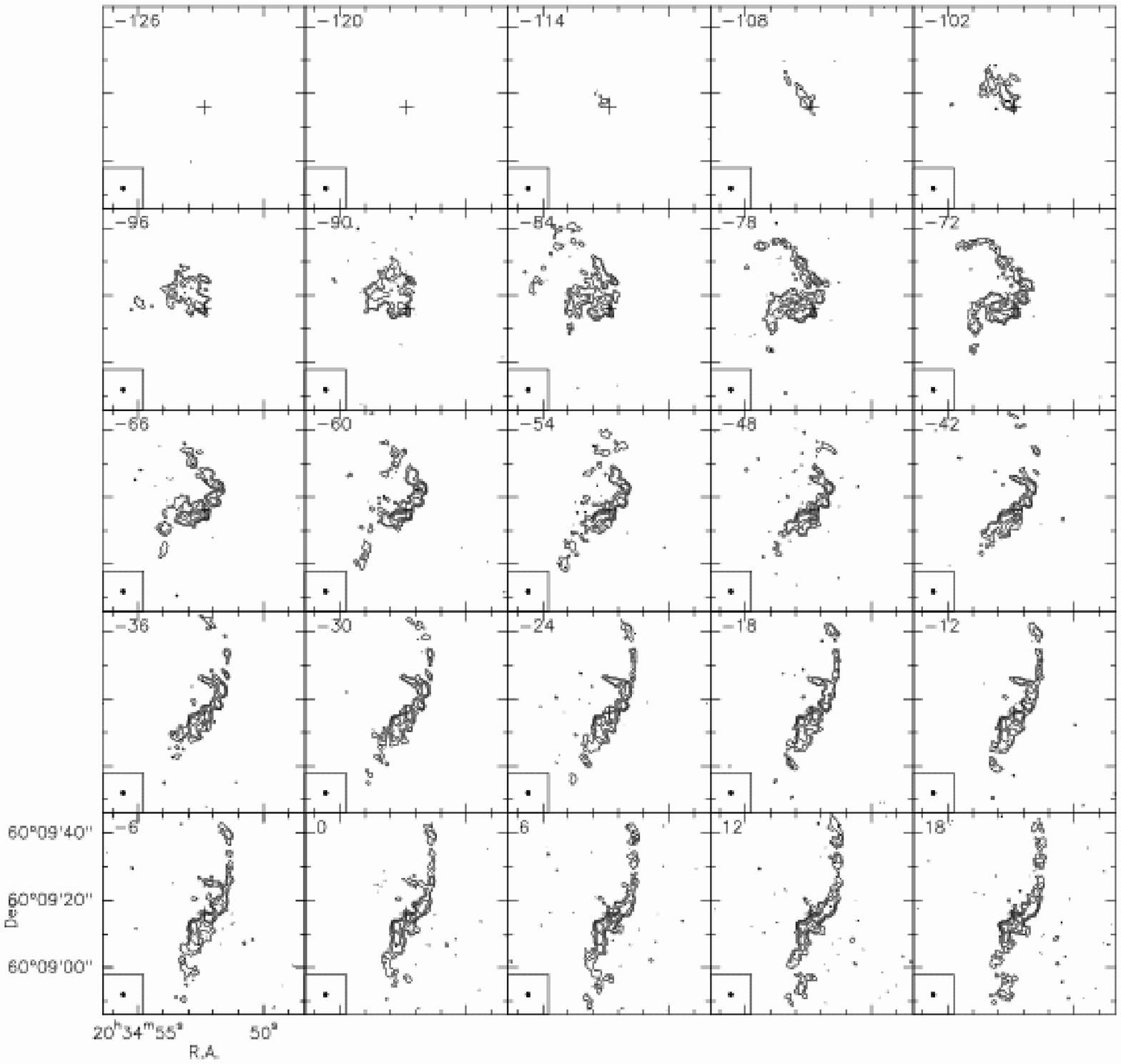}
\end{figure}
\begin{figure}
\includegraphics[angle=0,scale=.6]{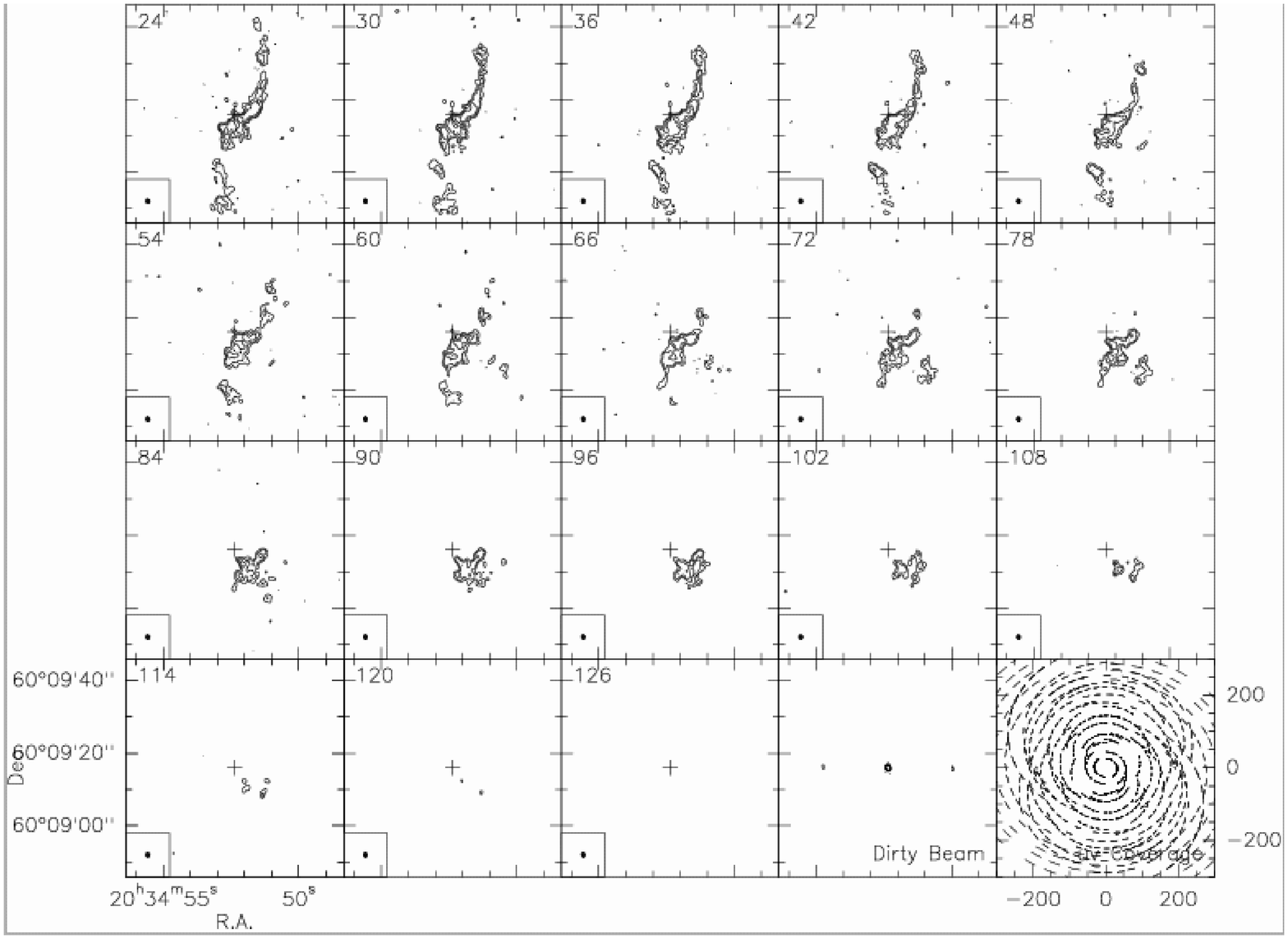}
\caption{Channel maps of the \coone\ line. The channels are $6\kms$ wide, 
and contours are plotted at $5,10,20,40\sigma$ with $1\sigma$ = 4 mJy/beam.
The velocity marking (top left corner) is relative to the observed central 
velocity of $v=65\kms$. The CLEAN beam of 1.25$\as \times$1.01$\as$ is 
shown in the bottom left corner of each channel. The dirty beam and the 
$uv$ coverage are shown in the last two channels.
\label{fig:10channels}
}
\end{figure}

\newpage
\begin{figure}
\includegraphics[angle=0,scale=.6]{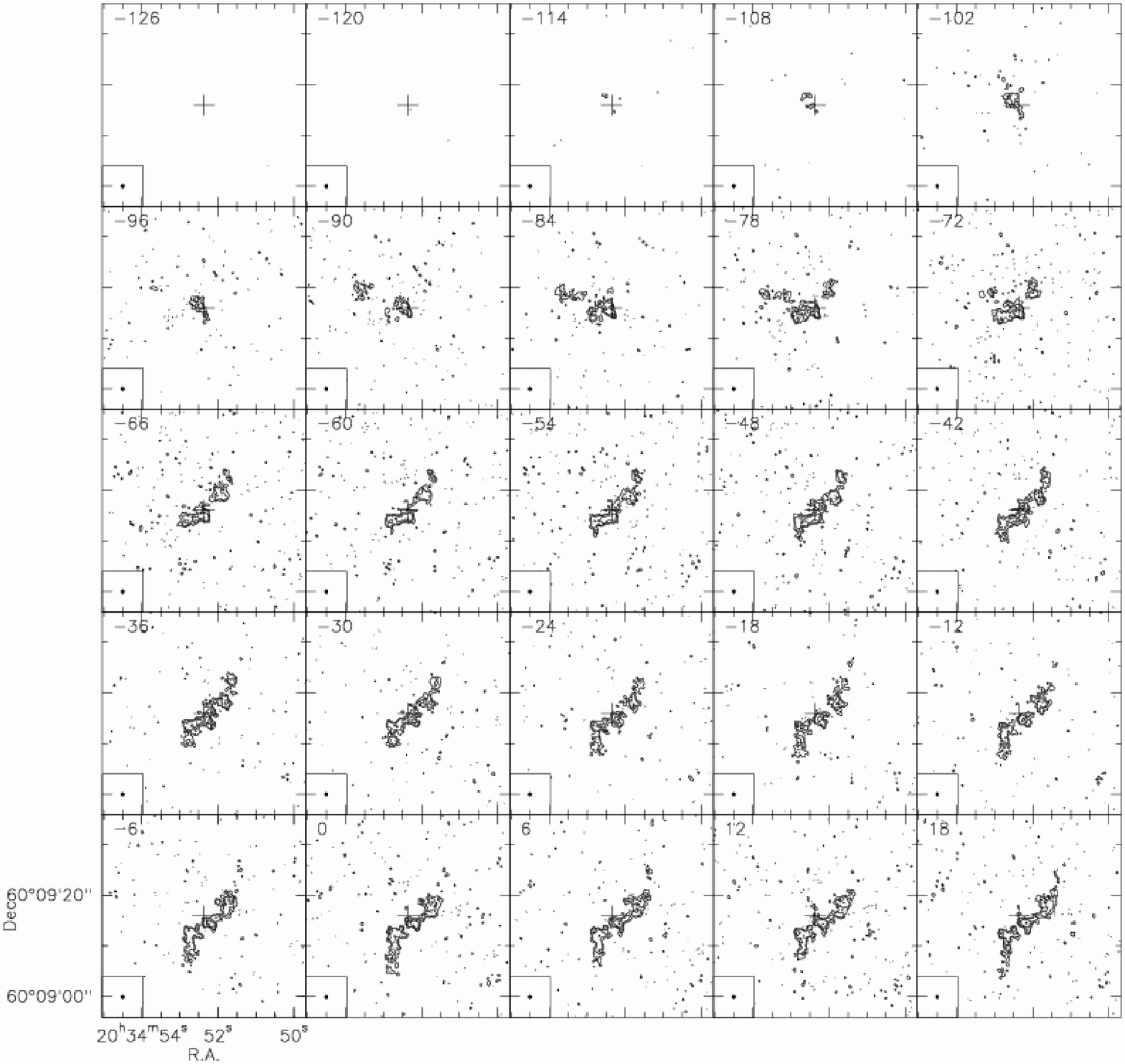}
\end{figure}
\begin{figure}
\includegraphics[angle=0,scale=.6]{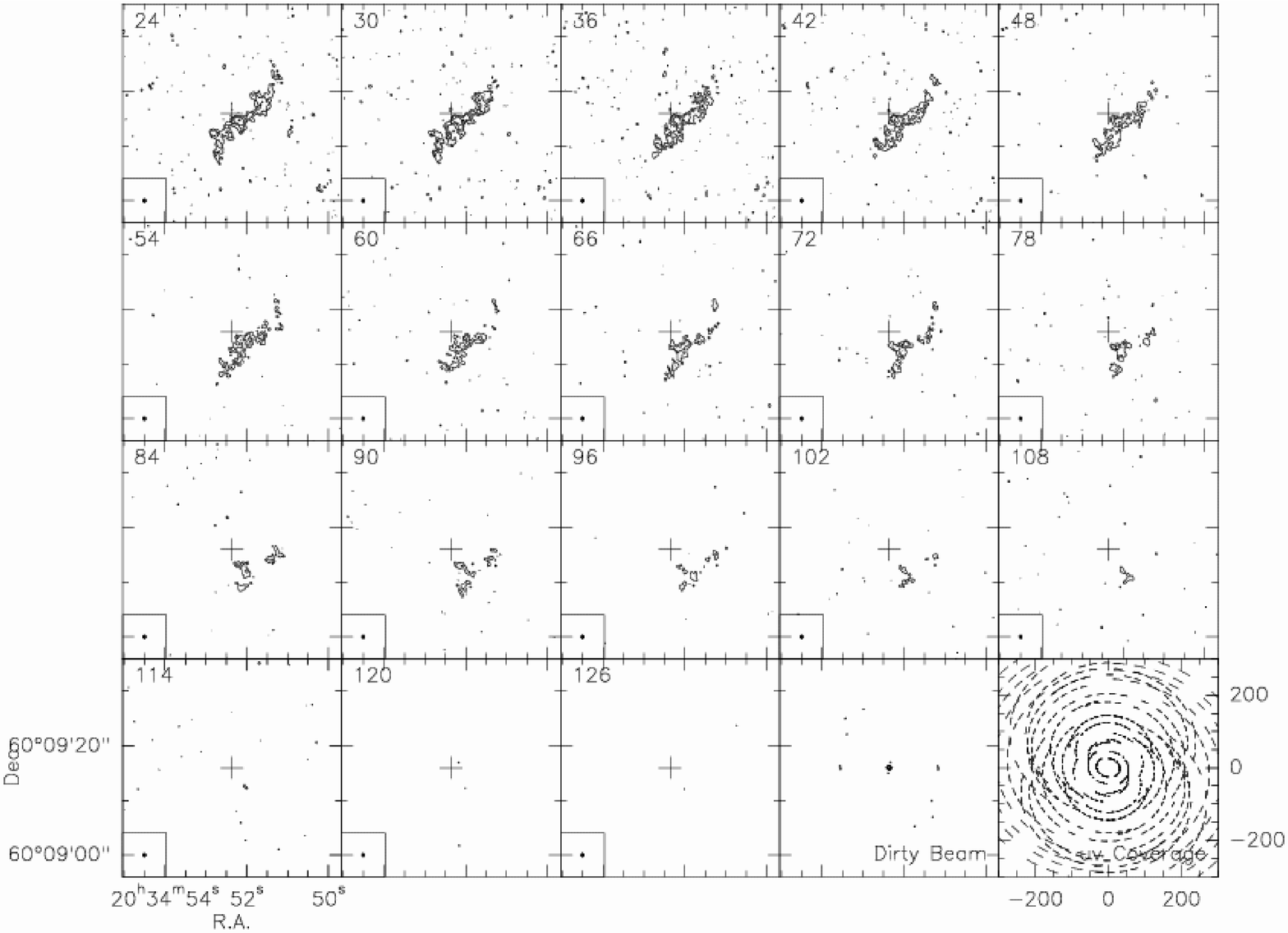}
\caption{Channel maps of the \cotwo\ line. The channels are $6\kms$ wide, 
and contours are plotted at $3,6,12,24\sigma$ with $1\sigma$ = 12 mJy/beam.
The velocity marking (top left corner) is relative to the observed central 
velocity of $v=65\kms$. The CLEAN beam of 0.58$\as \times$0.48$\as$ is 
shown in the bottom left corner of each channel. The dirty beam and the 
$uv$ coverage are shown in the last two channels.
\label{fig:21channels}
}
\end{figure}

\clearpage
\begin{figure}
\includegraphics[angle=-90,scale=.85]{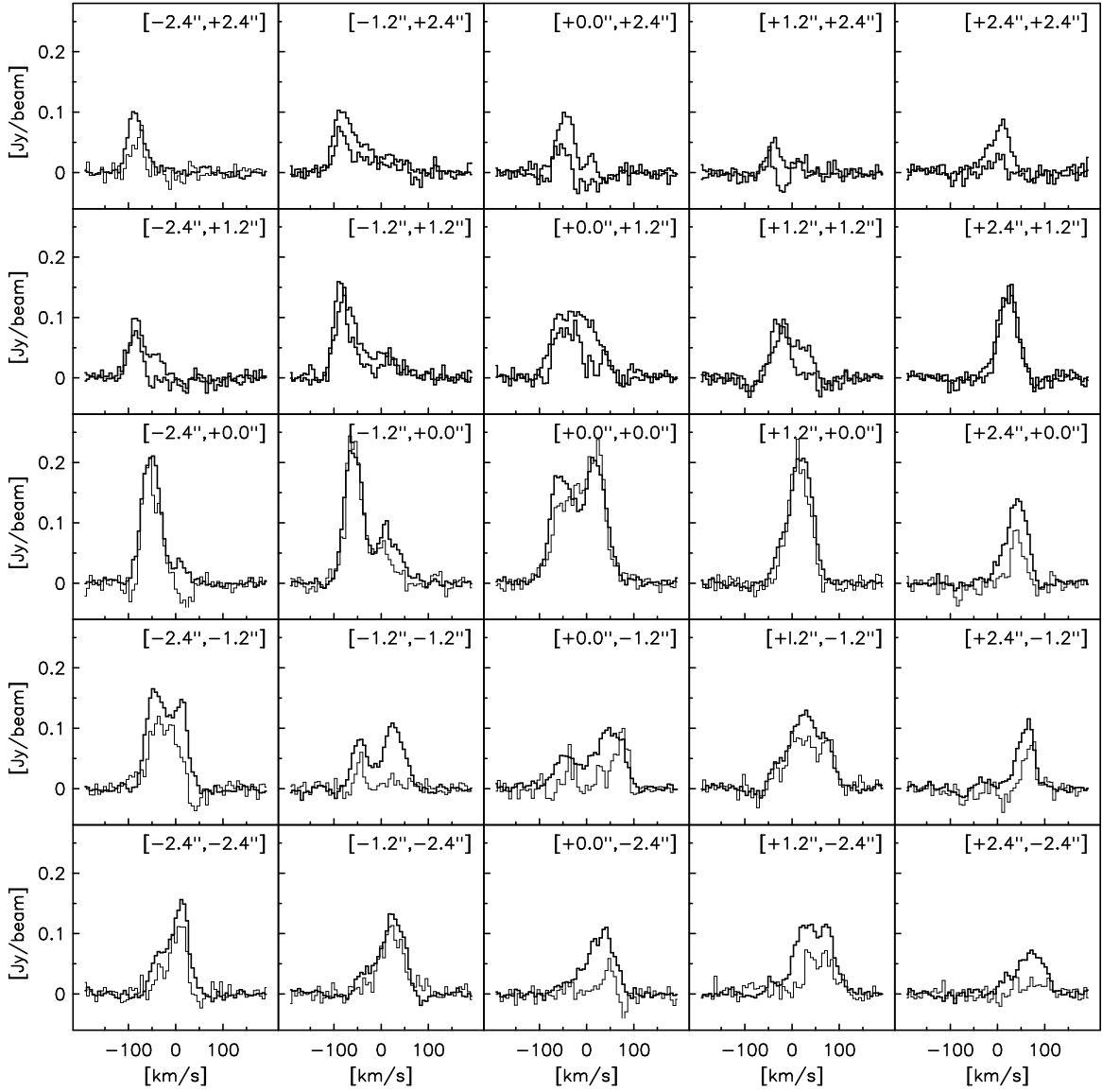}
\caption{Spectra of the \coone\ ({\it heavy solid line}) and 
  \cotwo\ ({\it thin solid line}) line emission extracted with in the
  central $4.8\as\,\times4.8\as$ around the dynamical center (see \S\,
\ref{subsec:kinematics}). The spectra have not been corrected for the 
different beam sizes.
\label{fig:spec}
}
\end{figure}

\clearpage
\begin{figure}
\includegraphics[angle=0,scale=.70]{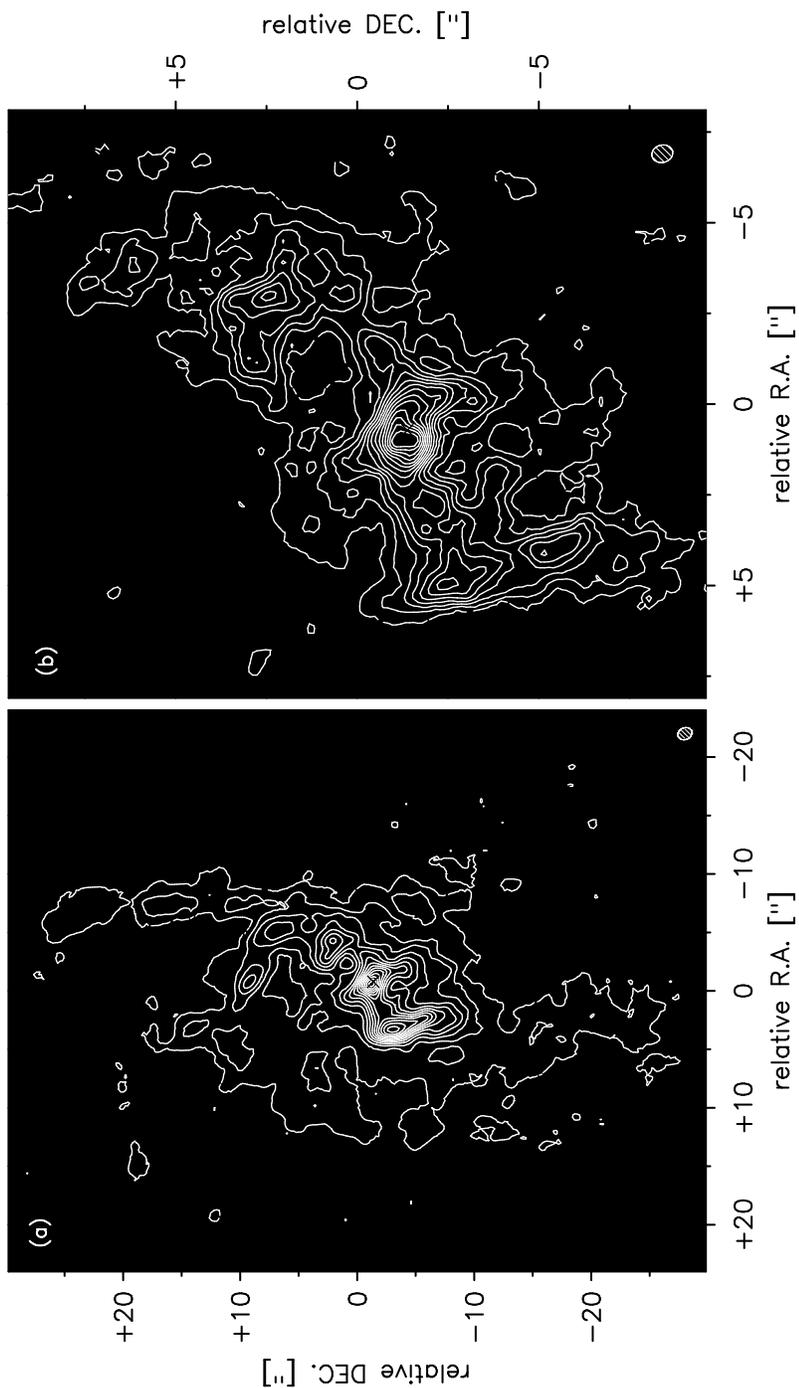}
\caption{Intensity maps of the \coone\ ({\it a}) and \cotwo\ ({\it b}) 
lines. The coordinates are relative to the phase center of the observations.
The dynamical center inferred from the kinematics (see \S\,
\ref{subsec:kinematics}) is marked with a cross in both panels. The  
contour levels are 0.60, 2.28, 3.96, ... $\jy\kms$ and 1.26, 2.94, 
4.62, ... $\jy\kms$ for the \coone\ and \cotwo\ emission.
\label{fig:co-sum}
}
\end{figure}

\clearpage
\begin{figure}
\includegraphics[angle=0,scale=.80]{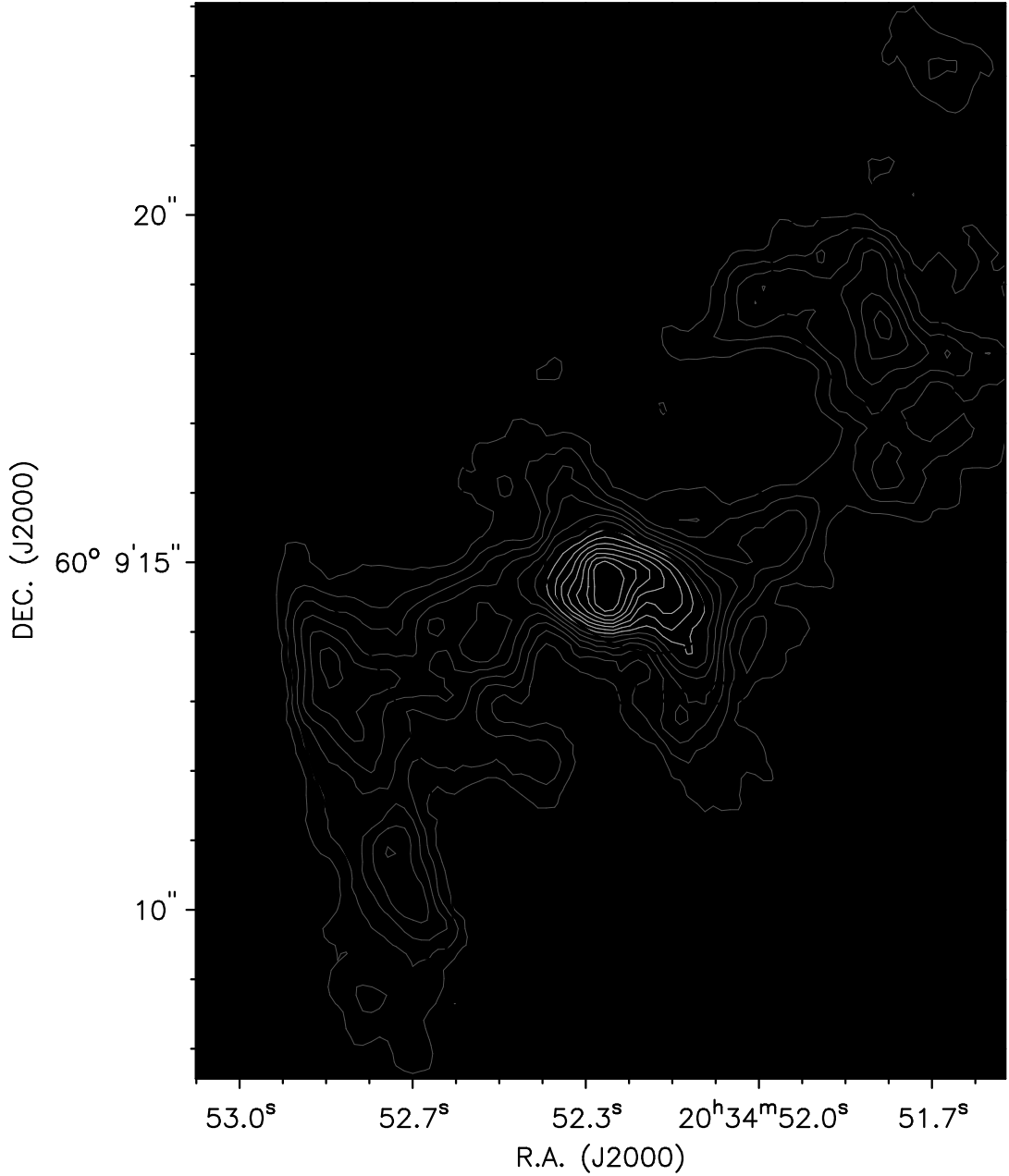}
\caption{\cotwo\ intensity map ({\it gray-scale, grey contours}) showing the nuclear 
 spiral structure.  The important features described in the text (see
 \S\,\ref{subsec:mass}) are indicated. The ellipses roughly outline
 the areas used to derived the \coone\ line flux and thus the
 molecular gas mass (see \S\,\ref{subsec:mass} and
 Tab.\,\ref{tab:COfluxes}). The CLEAN beam is shown in the bottom left
 corner.
\label{fig:feat}
}
\end{figure}

\clearpage
\begin{figure}
\includegraphics[angle=0,scale=.70]{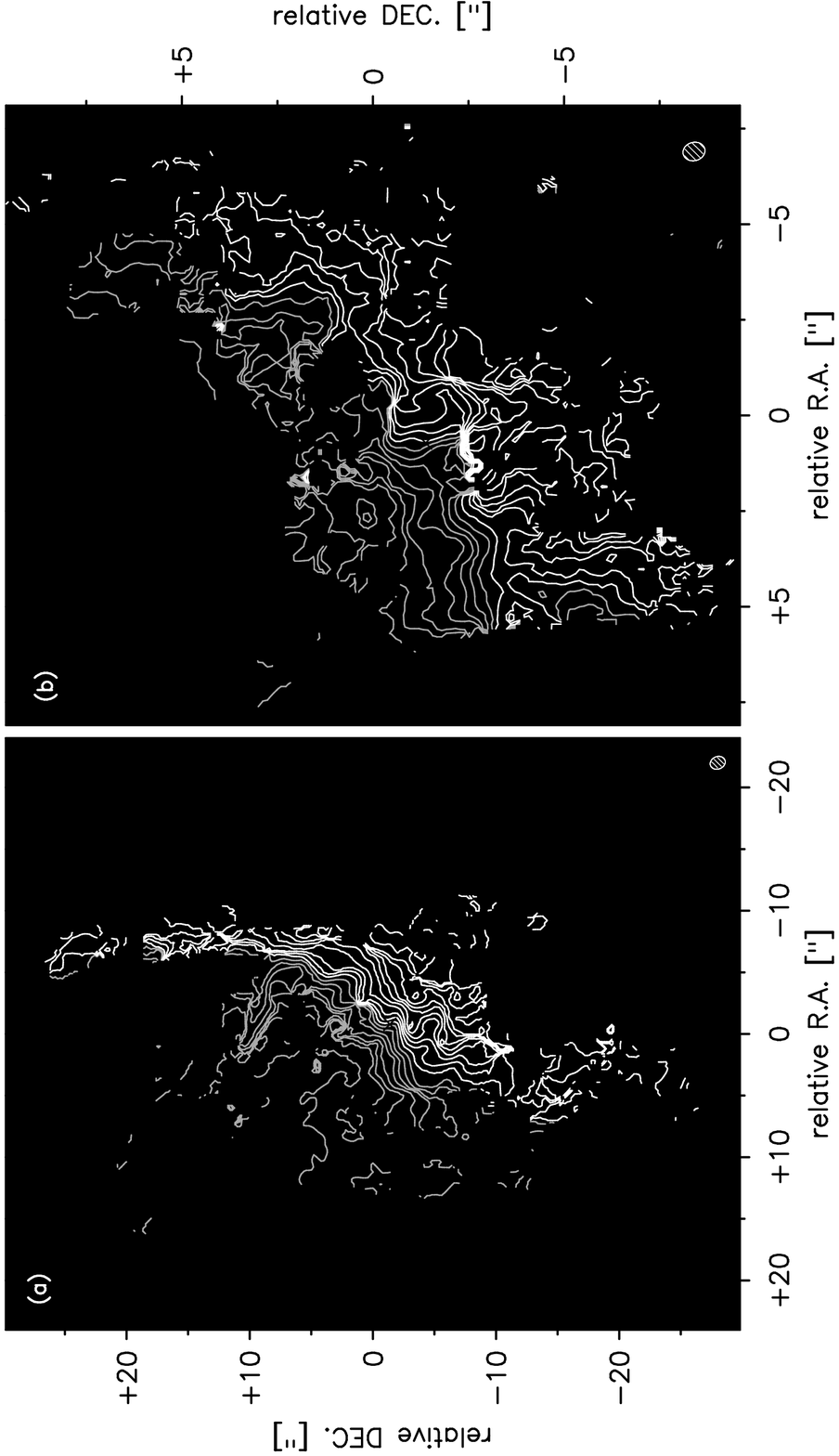}
\caption{Velocity fields of the \coone\ ({\it a}) and \cotwo\ ({\it b}) 
lines. The coordinates are relative to the phase center of the observations.
The dynamical center inferred from the kinematics (see \S\,
\ref{subsec:kinematics}) is marked with a cross in both panels. 
The iso-velocity contours are in steps of $10\kms$
starting from $+5\kms$ (positive; white contours) and $-5\kms$ 
(negative; grey contours) relative to the systemic velocity of
$v_{sys}(LSR)=50\kms$ (see \S \ref{subsec:kinematics}). 
\label{fig:co-vel}
}
\end{figure}

\clearpage
\begin{figure}
\includegraphics[angle=0,scale=.75]{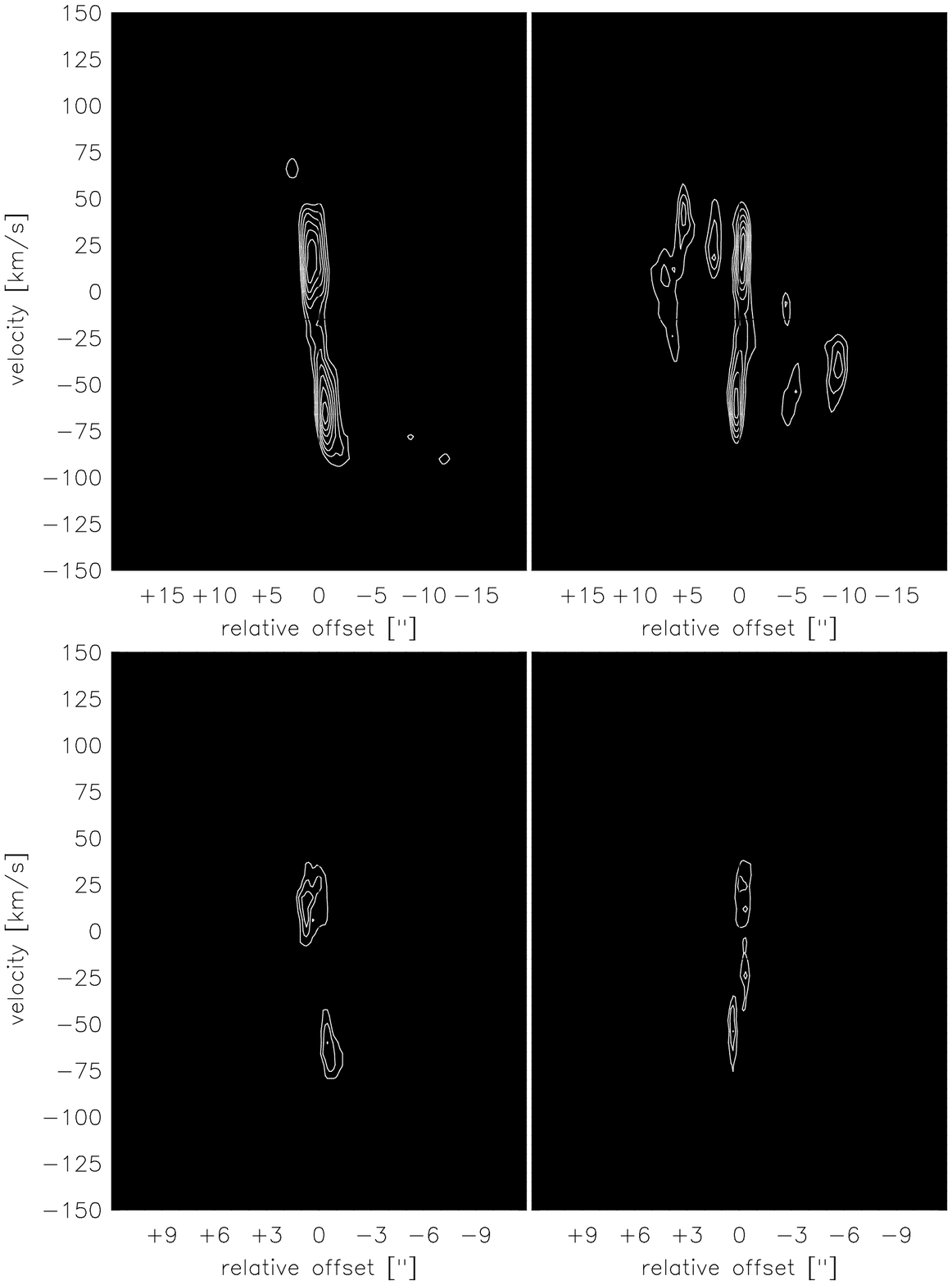}
\caption{Position-velocity (p-v) diagrams of the \coone\ ({\it top}) and 
  \cotwo\ ({\it bottom}) line emission along the major kinematic axes
  ({\it a, c}) and the minor kinematics axes ({\it b, d}). The
  contours start at 3$\sigma$=12mJy/beam and are in steps of 3$\sigma$
  till 24$\sigma$ when they continue in steps of 6$\sigma$ for the
  \coone\ p-v diagrams ({\it a, b}), and they start at
  3$\sigma$=30mJy/beam and are in steps of 3$\sigma$ for the \cotwo\
  ones ({\it b, d}). The solid horizontal line corresponds to the
  systemic velocity of $v_{sys}(LSR)=50\kms$.  
\label{fig:pv}
}
\end{figure}

\clearpage
\begin{figure}
\includegraphics[angle=0,scale=.70]{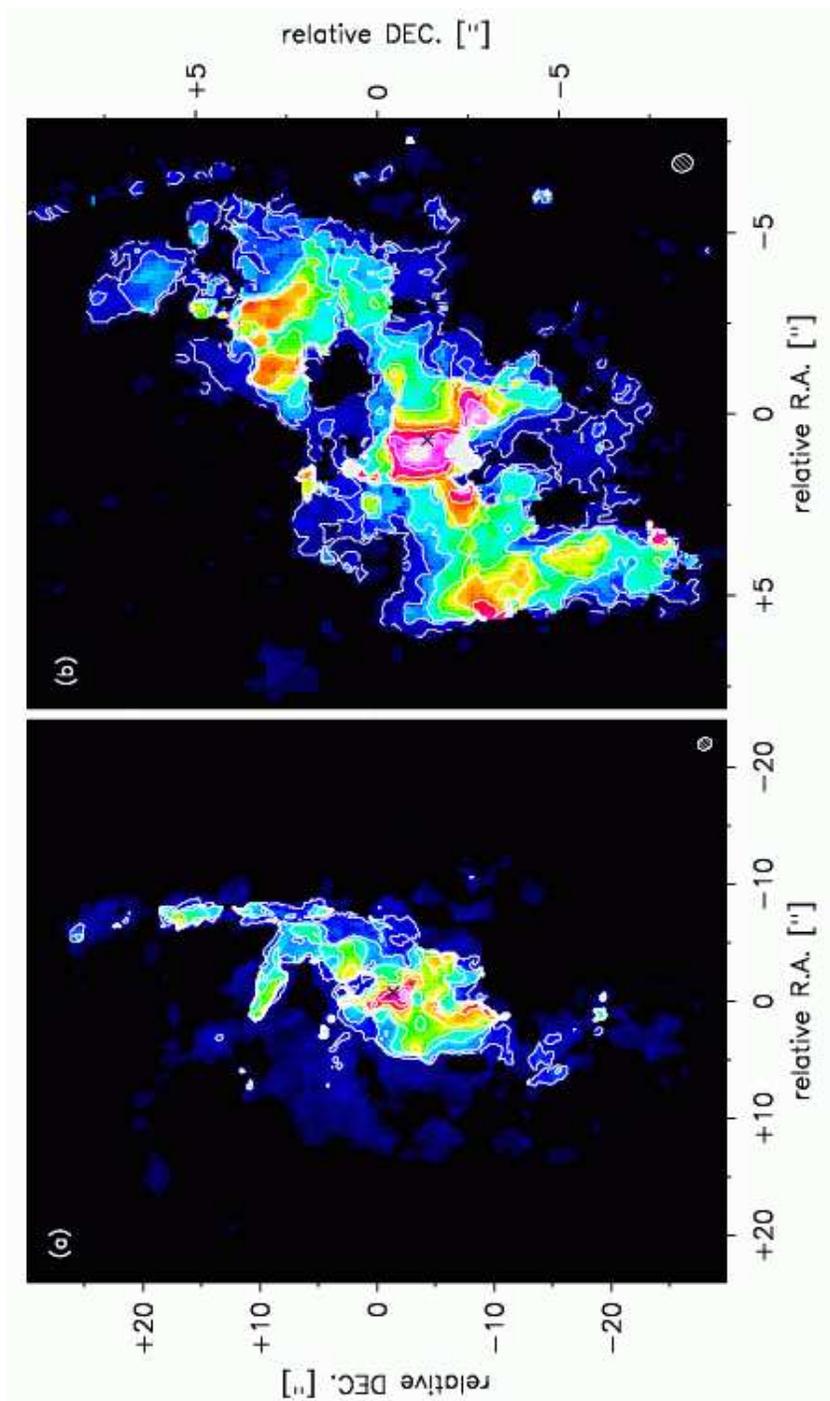}
\caption{Dispersion maps of the \coone\ ({\it a}) and \cotwo\ ({\it b}) 
lines. The coordinates are relative to the phase center of the observations.
The dynamical center inferred from the kinematics (see \S\,
\ref{subsec:kinematics}) is marked with a cross in both panels. The contours 
start at $10\kms$ and $5\kms$ for the \coone\ and \cotwo\ line
with steps of $5\kms$.
\label{fig:co-disp}
}
\end{figure}

\clearpage
\begin{figure}
\includegraphics[angle=0,scale=.75]{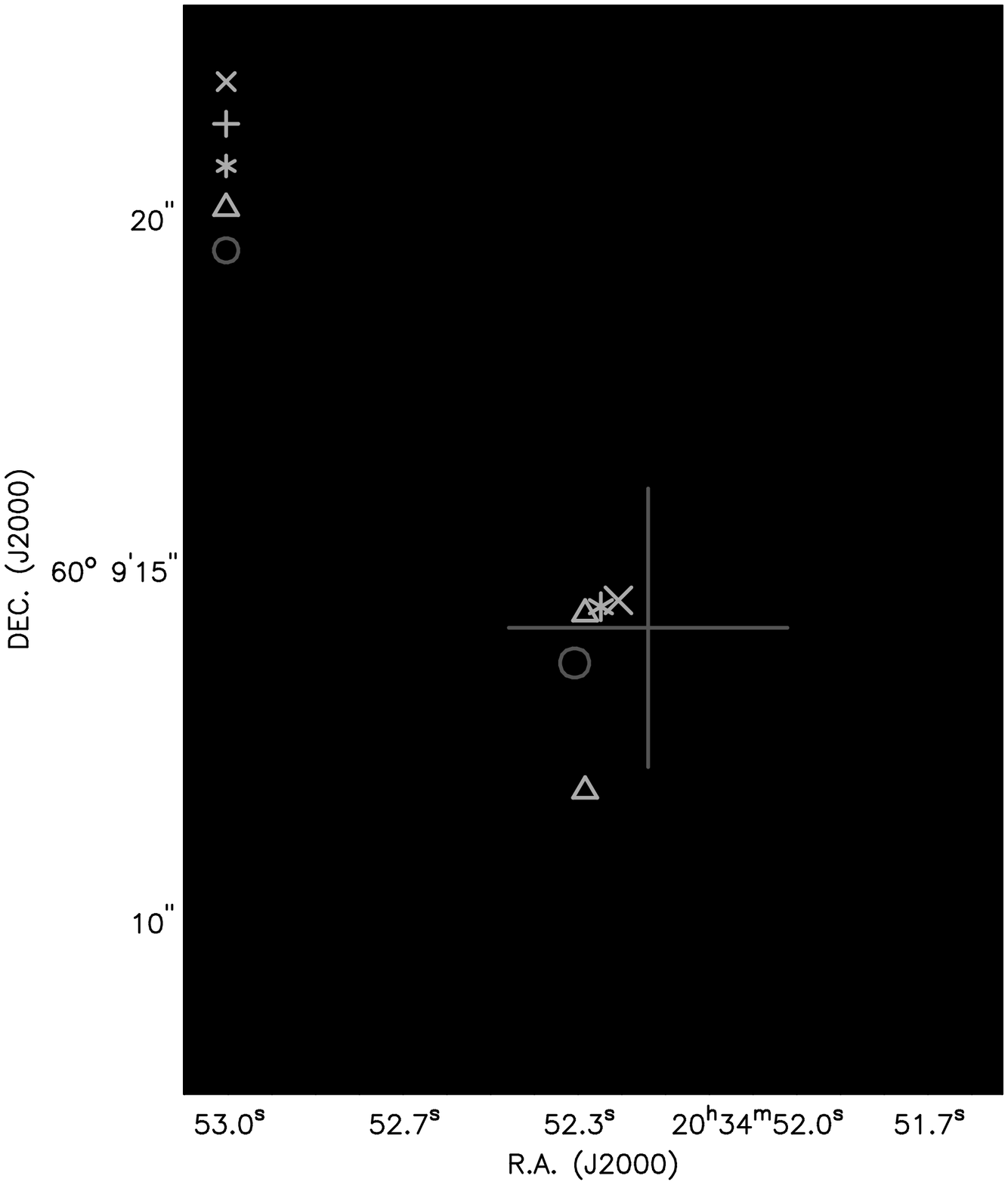}
\caption{Location of the dynamical center as derived from our data
  (\cotwo\ intensity map in grey-scale),
  compared to that derived from observations made with the OVRO 
  mm-interferometer \citep{mei04}. Also indicated are the positions of 
  the 2\,cm radio continuum peak
  (Turner \& Hurt 1983) and two nuclear X-ray sources identified by
  \cite{hol03}. In addition, the position of the 2MASS center from
  \cite{mei04} is shown. The size of the symbol represents the positional
  uncertainty, except for the X-ray sources. The solid lines indicate the 
  orientation
  of the position-velocity cuts presented in Fig. \ref{fig:pvbar}.
  The CLEAN beam is shown in the bottom left corner.
\label{fig:center}
}
\end{figure}

\clearpage
\begin{figure}
\includegraphics[angle=0,scale=.70]{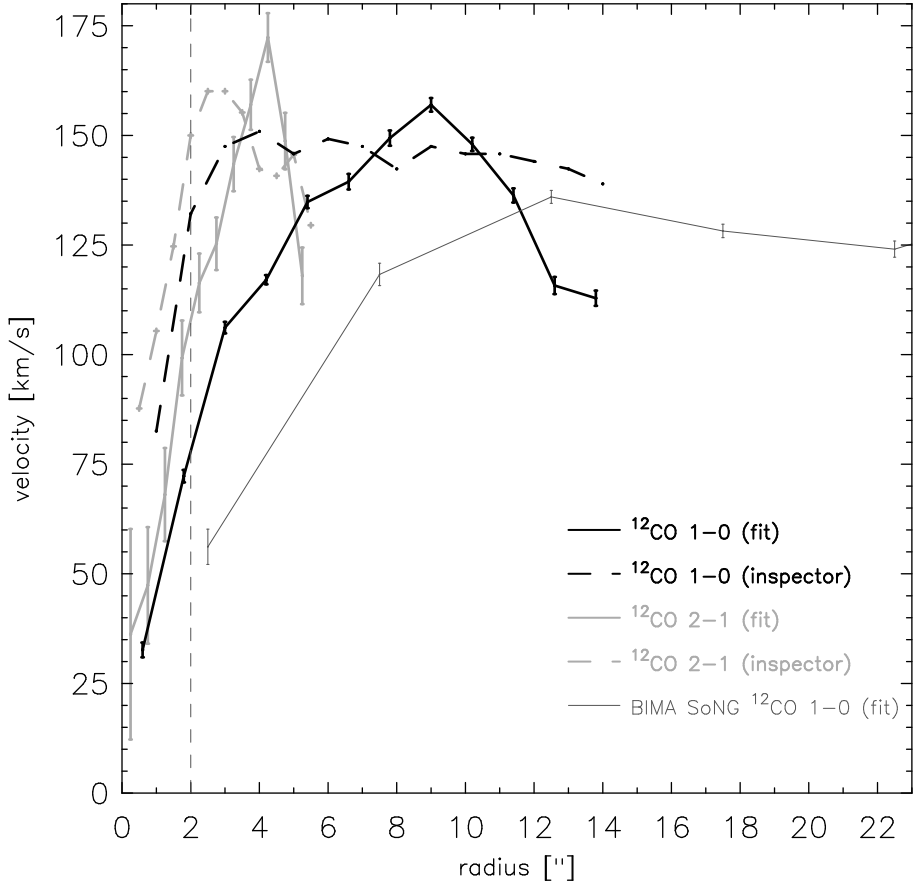}
\caption{Rotation curves of the molecular gas within a radius of 
$\sim\,500\pc$ from the nucleus of NGC\,6946. The derived deprojected
velocities for the \coone\ and \cotwo\ line from the PdBI and BIMA
data are shown. The steepening of the gradient within the first 6$\as$
is due to the higher angular resolution of the \cotwo\ data and the
fact that the 'INSPECTOR' derived rotation curves are less affected by
beam smearing (see \S\,\ref{subsec:rot} for details). The error bars
for the 'ROTCUR' derived rotation curves represent the uncertainties
from the least-square fit.  The \coone\ rotation curve of the BIMA
SoNG data shows that the apparent drop in the PdBI CO rotation curves
is an artifact due to insufficient sampling at larger radii. The
dashed line marks the radius of 2$\as$ used to derive the dynamical
mass in \S\,\ref{subsec:mass}.
\label{fig:rot}
}
\end{figure}

\clearpage
\begin{figure}
\includegraphics[angle=0,scale=.80]{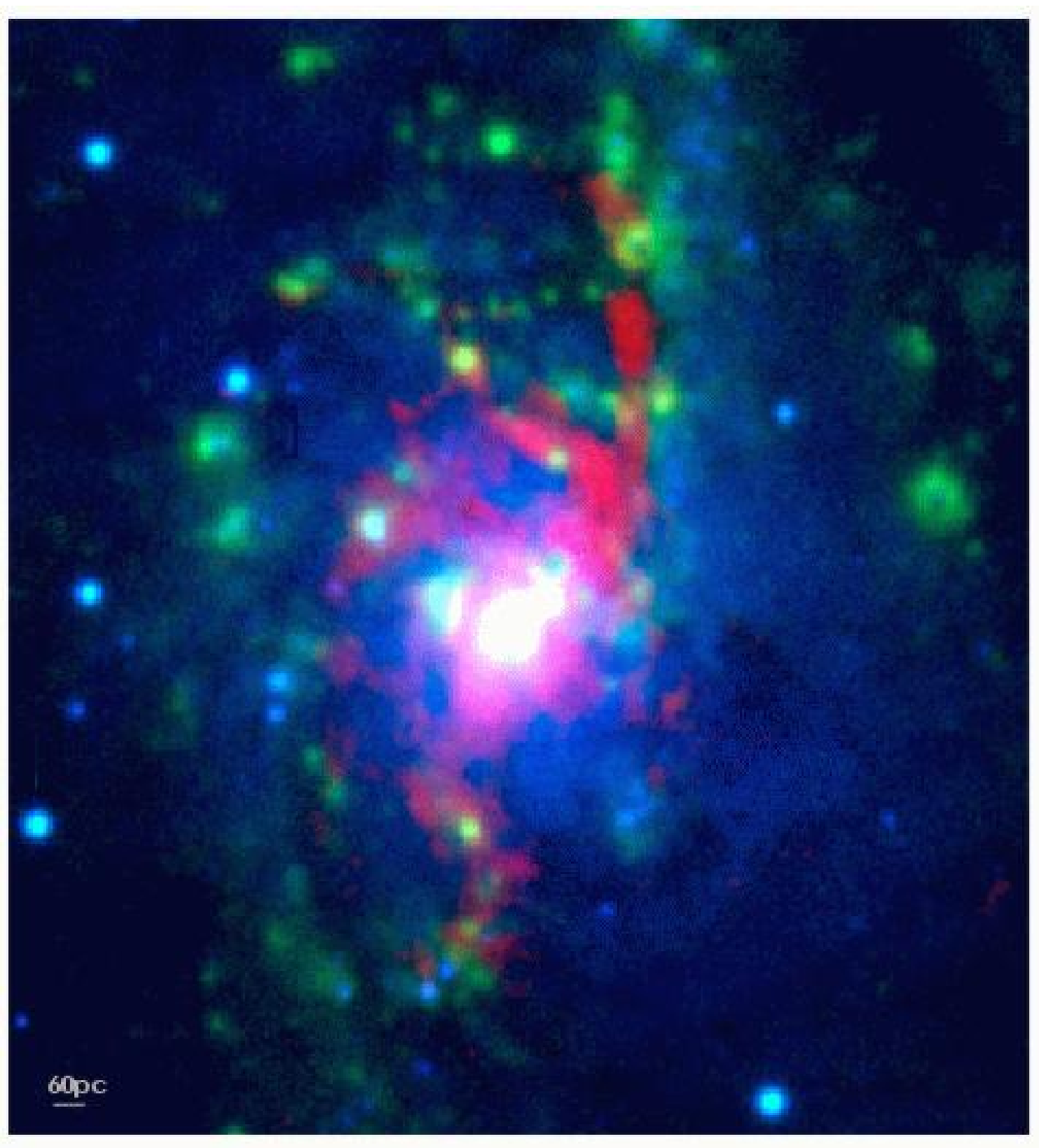}
\caption{Three color composite image of the central arcminute in 
NGC\,6946 showing the \coone\ line emission ({\it red}), the \ha\ line
emission ({\it green}) and the I band continuum ({\it blue}) from the NOT
telescope \citep{lar99}. The offset between the \ha\ and \coone\
line emission along the northern straight gas lane is very reminiscent
of what has been commonly observed for gas and dust lanes along the
leading side of large-scale stellar bars \citep{she02}.
\label{fig:rgb}
}
\end{figure}

\clearpage
\begin{figure}
\includegraphics[angle=0,scale=.80]{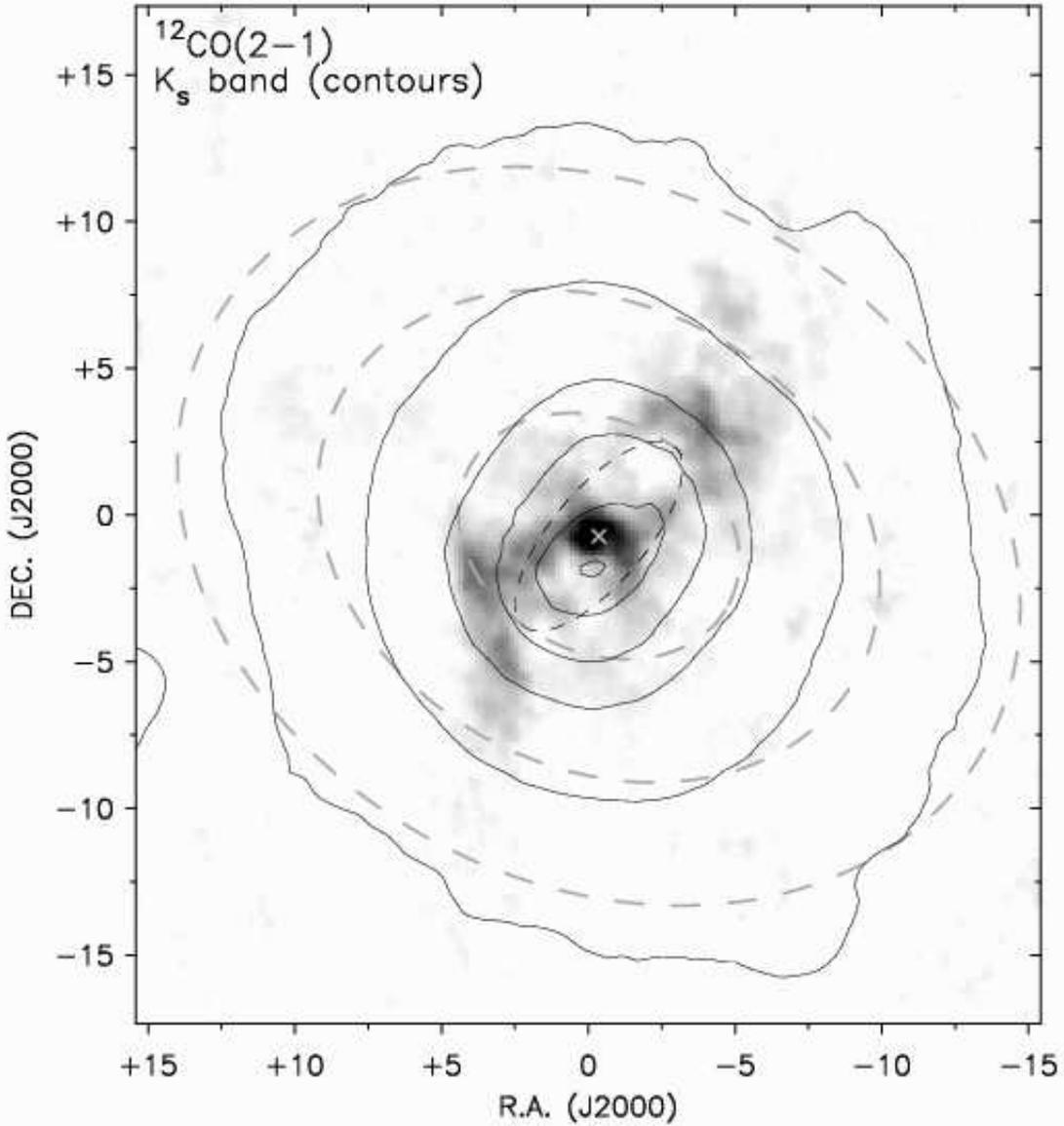}
\caption{Overlay of the K$_s$ band isophotes from the \cite{kna03} image 
onto the \cotwo\ intensity map (grey-scale). The extent of the stellar
bar ($4\as$) as determined by \cite{elm98} is shown as a dashed
ellipse. The grey dashed ellipses describe circles within the plane
of the galaxy with radii of $5\as$, $10\as$, and $15\as$. The cross
marks the location of the dynamical center as described in
\S\,\ref{subsec:rot}.  For further explanation see \S\,\ref{subsec:bar}.
\label{fig:bar}
}
\end{figure}

\clearpage
\begin{figure}
\includegraphics[angle=-90,scale=.65]{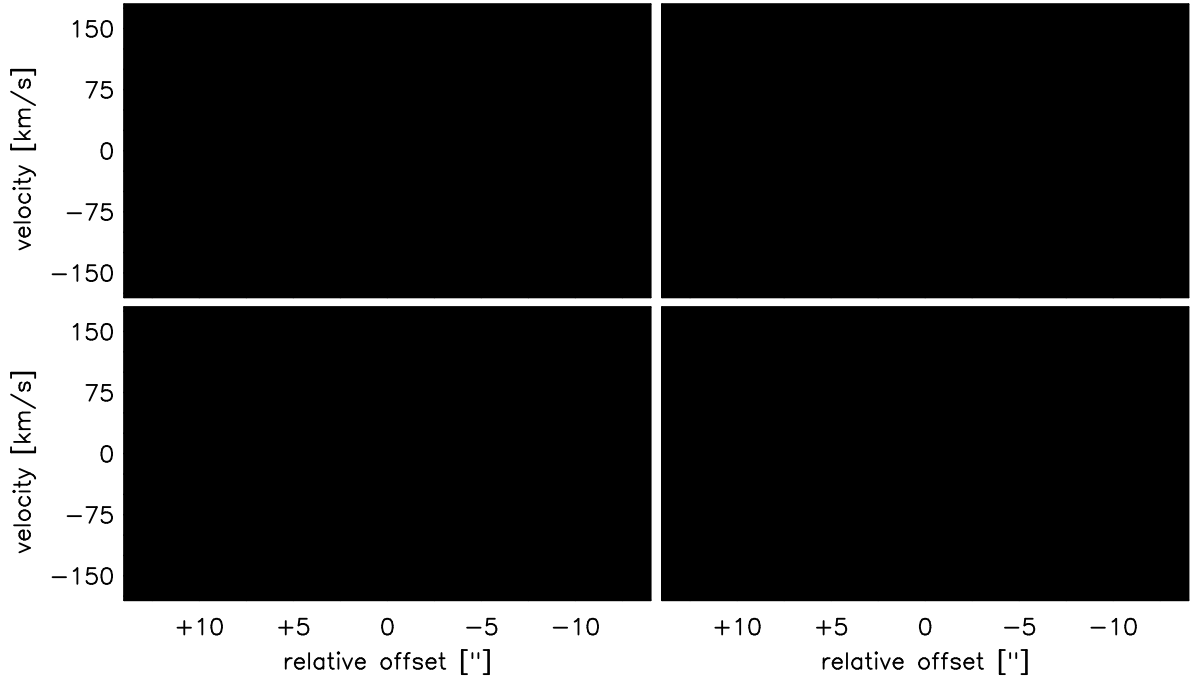}
\caption{\coone\ position-velocity 
  diagrams for two positions perpendicular to the southern ({\it
  right}) and northern ({\it left}) straight gas lanes (indicated in
  Fig. \ref{fig:center}). (See text for more details.)
\label{fig:pvbar}
}
\end{figure}

\clearpage
\begin{figure}
\includegraphics[angle=0,scale=.60]{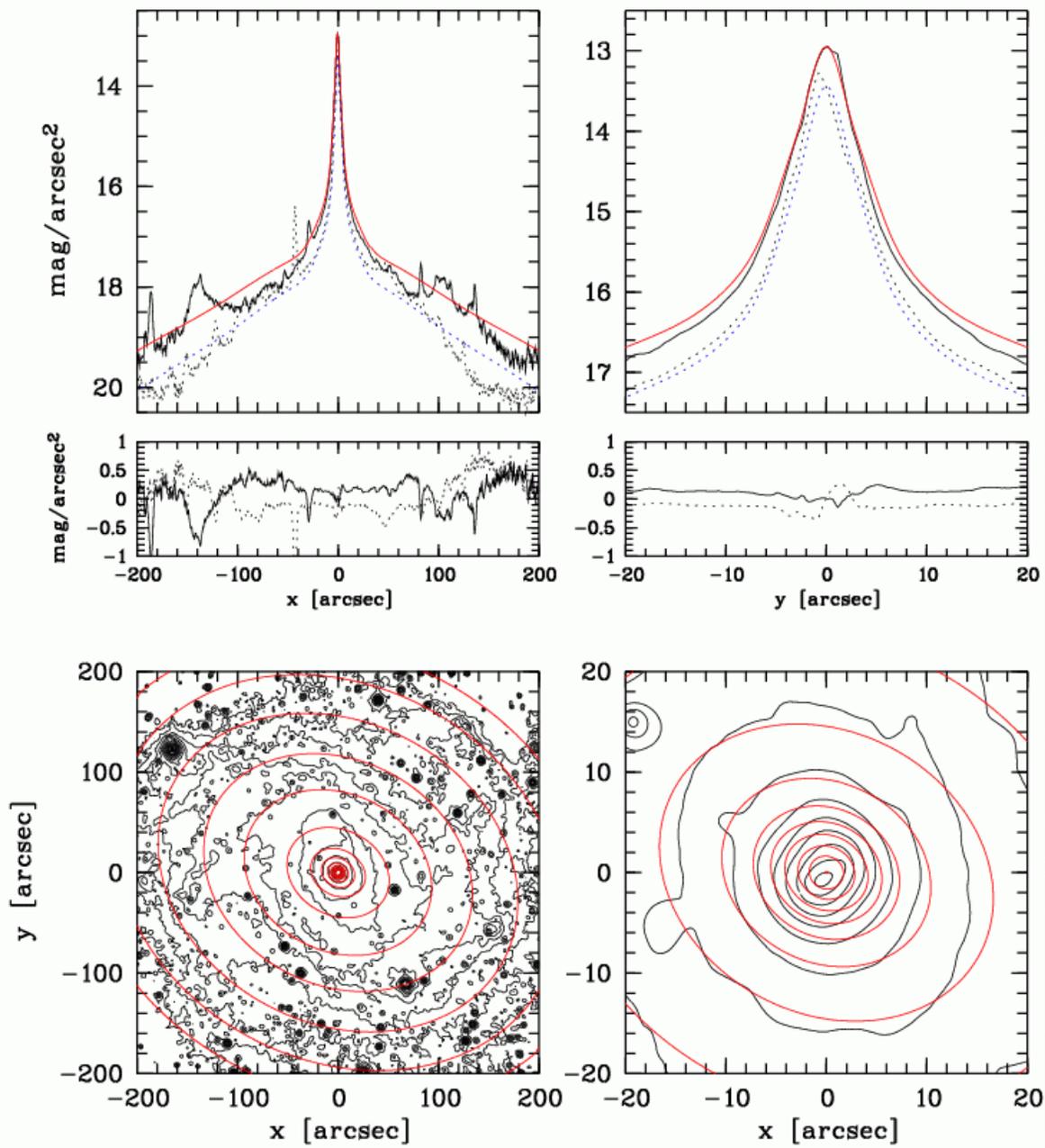}
\caption{Comparison between the $K_s$ band image and an axisymmetric
  luminosity model built using the MGE formalism (see text for
  details). {\it Top}: North-South (solid: $K_s$ band, red: MGE model)
  and East-West (dotted: $K_s$ band, blue: MGE model) cuts for the
  central $200\as$ ({\it left}) and $20\as$({\it right}). {\it
  Middle:} Residuals between the model and the data. Larger
  discrepancies are due to the large-scale bar ({\it left}) and the
  depression of emission due to the high extinction in the nucleus
  ({\it right}) which are not taking into account in the axisymmetric
  model. {\it Bottom}: isophotes of the $K_s$ band images (black
  contours) and of the MGE model (red contours).  }
\label{fig:massmodel}
\end{figure}

\clearpage
\begin{figure}
\includegraphics[angle=0,scale=.90]{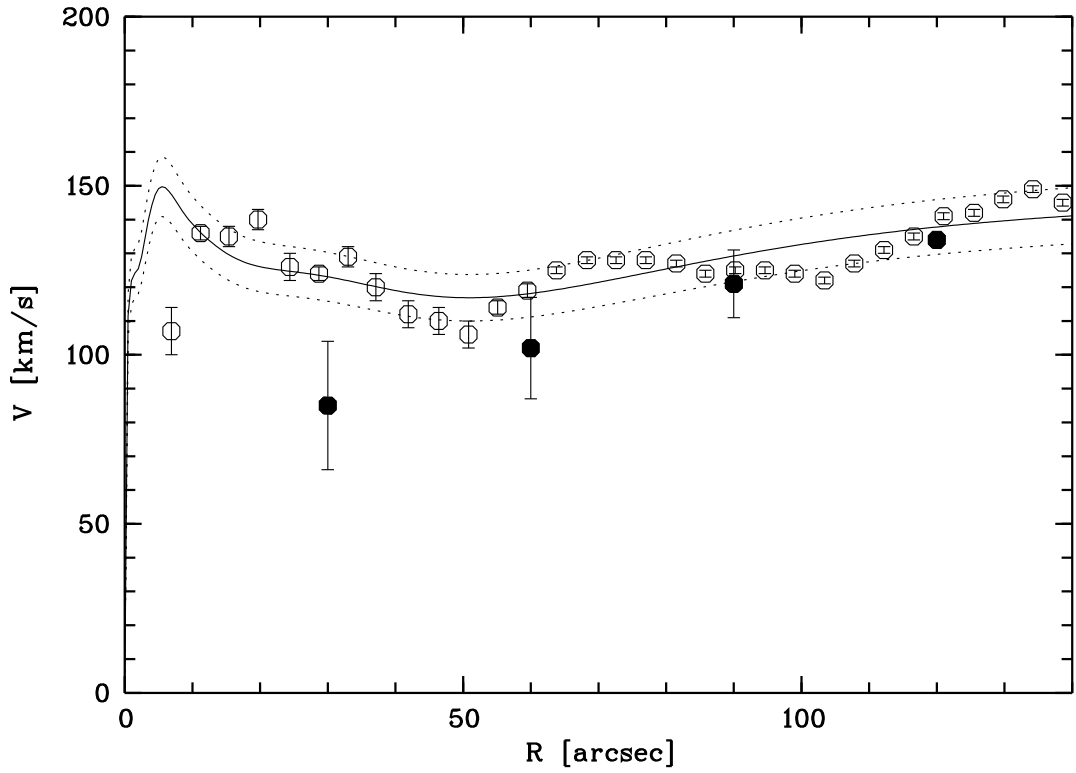}
\caption{Comparison between the observed (deprojected) HI ({\it 
filled circles}) and \ha\ ({\it open circles}) rotation curves in
the central 140\arcsec of NGC~6946, and the circular velocity profile
derived from the MGE model for $\rm M/L_K$ of 0.72 ({\it solid lines}),
0.64 and 0.81 ({\it dashed lines}).}
\label{fig:Vcmodel}
\end{figure}

\clearpage
\begin{figure}
\includegraphics[angle=0,scale=.75]{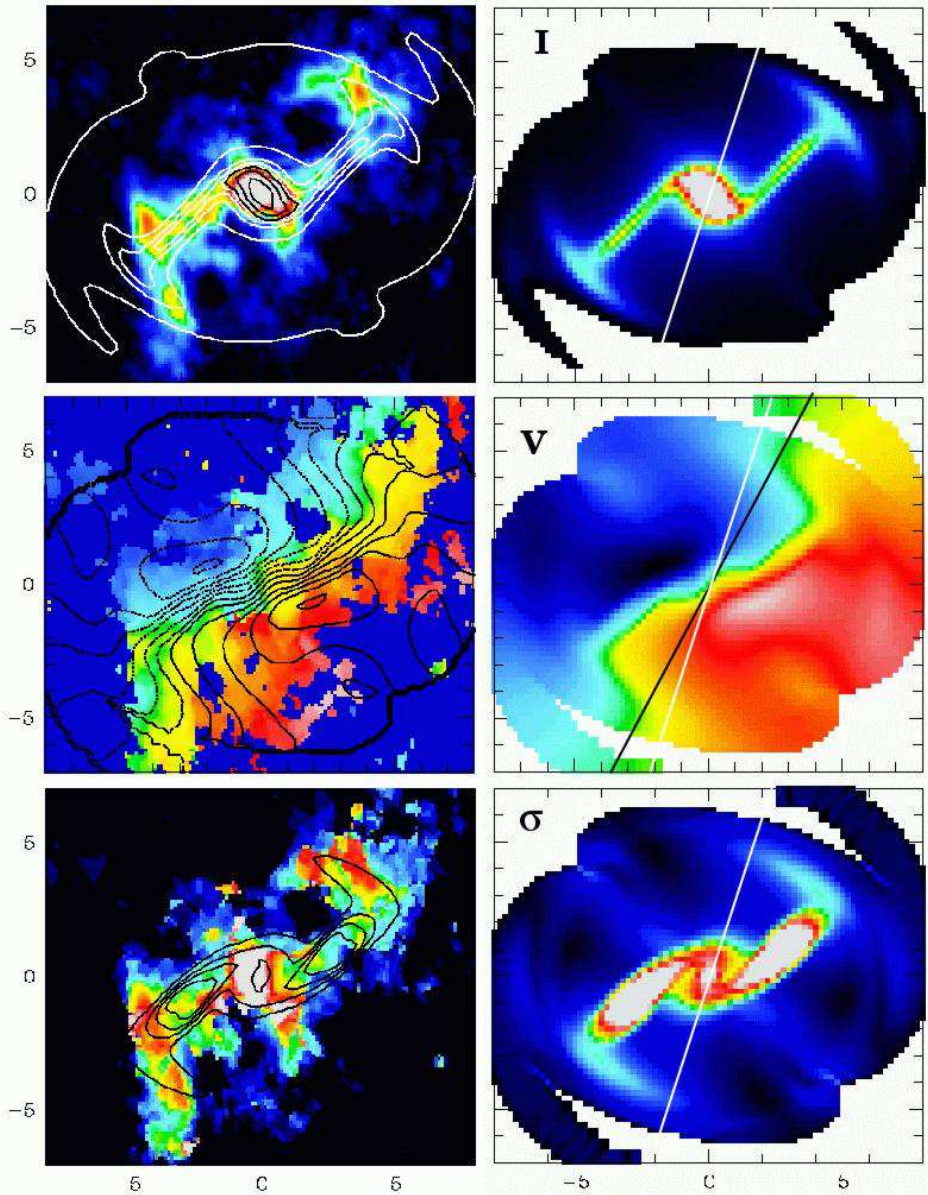}
\caption{Comparison between the intensity maps ({\it top}), the
  velocity fields ({\it middle}) and the velocity dispersion maps
  ({\it bottom}) of the gas component of the barred MGE model ({\it
  right:} color, {\it left:} contours) and the observed \cotwo\ line
  emission ({\it left:} color). In the {\it right} panels, we indicate
  the apparent position angle of the bar in the model (white dashed
  line) and the line-of-nodes of the unperturbed axisymmetric
  potential (black dotted line). See text for details.  }
\label{fig:kinmodel}
\end{figure}

\clearpage
\begin{figure}
\includegraphics[angle=0,scale=.70]{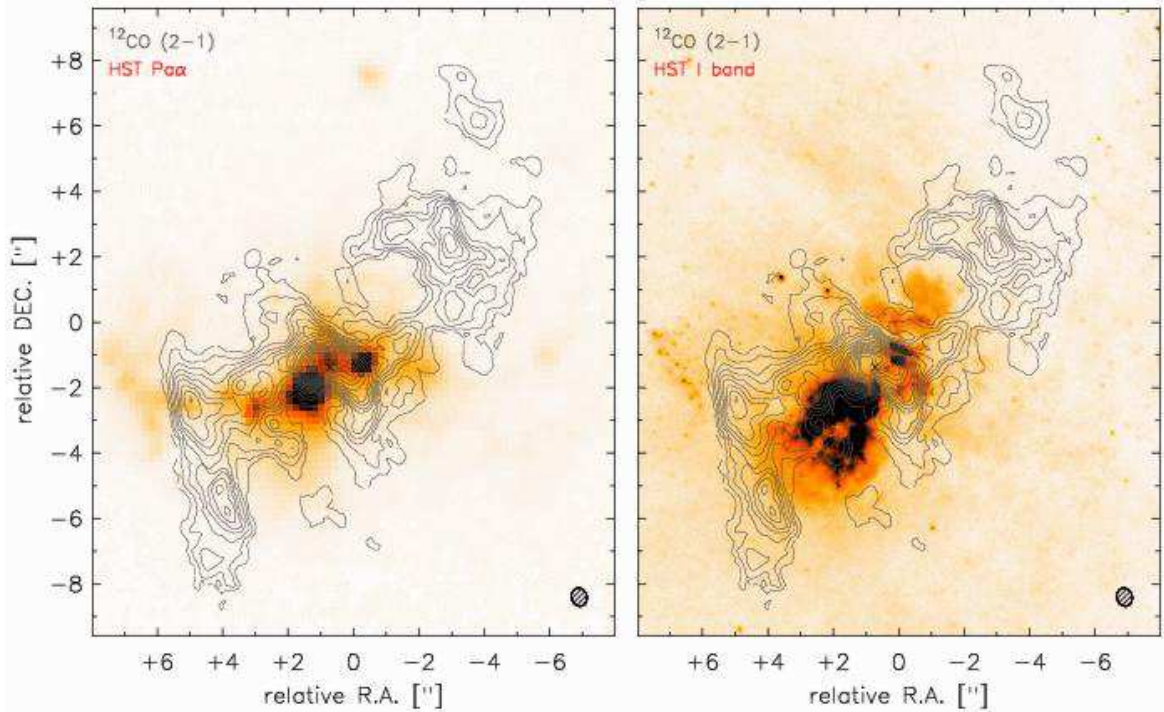}
\caption{Comparison of the CO distribution to HST maps of the stellar light
  as traced by the I-band continuum ({\it left}) and the current star
  formation as traced by the emission of the \paa\ recombination line
  ({\it right}). In both panels, the cross marks the location of the
  dynamical center as discussed in \S\,\ref{subsec:kinematics}. The
  dynamical center (and the highest CO luminosity and likely density)
  is located in a region of extreme extinction.
\label{fig:hstcomp}
}
\end{figure}

\end{document}